\documentclass[a4paper,11pt]{article}
             
\usepackage{jheppub} 
\usepackage[T1]{fontenc} 
\usepackage{amsmath,amssymb,xfrac,framed,verbatim,amsthm}
\usepackage{mathrsfs,accents,bbm,bbold} 
\usepackage{simplewick}
\usepackage{slashed,mathtools} 
\usepackage{sectsty} 
\usepackage{graphicx,enumerate,bm} 
\graphicspath{{figures/}}
\usepackage{verbatim}
\usepackage{comment}
\usepackage{array}
\usepackage{longtable}
\usepackage{float}
\restylefloat{table}
\usepackage{soul} 
\usepackage{cancel}
\usepackage{amsmath}
\allowdisplaybreaks

\newcommand{\sgn}{\text{sgn}}
\renewcommand{\i}{\text{i}}

\usepackage[multiple]{footmisc}

\title{\boldmath The Characteristics of the Z-Boson in Chern-Simons Matter Theory}

\author{Amiya Mishra}
\emailAdd{mishramiya@gmail.com}

\affiliation[a]{Shing-Tung Yau Center and School of Physics, Southeast University,\\ 
	Yifu Architecture Building, No.2 Sipailou, Xuanwu District, Nanjing, Jiangsu, 210096, China}
\affiliation[b]{International Centre for Theoretical Sciences-TIFR,\\ 
	Shivakote, Hesaraghatta Hobli, Bangalore 560089, India}

\usepackage{feynmp}

\usepackage{tikz}
\usepackage[compat=1.1.0]{tikz-feynman}

\abstract{We consider the Chern-Simons theory coupled to massive fundamental matter in three spacetime dimensions, focusing on the Higgsed phase of the bosonic matter in the large $N$ limit. To study the characteristics of the $Z$-boson in the Higgsed phase and address an apparent puzzle regarding its fermionic analogue, we employ the 3d Bose-Fermi duality to derive the exact large $N$ result for the two-point function of the $Z$-boson. We study the spectral representation of the $Z$-boson and find that, in the large $N$ limit, it behaves as an unstable resonance with a non-perturbatively large lifetime at weak coupling.}

\begin{document}
\maketitle

\section{Introduction}
 In three-dimensional Chern-Simons matter theories, a bosonization duality has been extensively tested through various calculations over the past decade \cite{Giombi:2011kc, Aharony:2011jz, Maldacena:2011jn, Maldacena:2012sf,
 	Chang:2012kt, Jain:2012qi, Aharony:2012nh, Yokoyama:2012fa,
 	GurAri:2012is, Aharony:2012ns, Jain:2013py, Takimi:2013zca,
 	Jain:2013gza, Yokoyama:2013pxa, Bardeen:2014paa, Jain:2014nza,
 	Bardeen:2014qua, Gurucharan:2014cva, Dandekar:2014era,
 	Frishman:2014cma, Moshe:2014bja, Aharony:2015pla, Inbasekar:2015tsa,
 	Bedhotiya:2015uga, Gur-Ari:2015pca, Minwalla:2015sca,
 	Radicevic:2015yla, Geracie:2015drf, Aharony:2015mjs,
 	Yokoyama:2016sbx, Gur-Ari:2016xff, Karch:2016sxi, Murugan:2016zal,
 	Seiberg:2016gmd, Giombi:2016ejx, Hsin:2016blu, Radicevic:2016wqn,
 	Karch:2016aux, Giombi:2016zwa, Wadia:2016zpd, Aharony:2016jvv,
 	Giombi:2017rhm, Benini:2017dus, Sezgin:2017jgm, Nosaka:2017ohr,
 	Komargodski:2017keh, Giombi:2017txg, Gaiotto:2017tne,
 	Jensen:2017dso, Jensen:2017xbs, Gomis:2017ixy, Inbasekar:2017ieo,
 	Inbasekar:2017sqp, Cordova:2017vab, Charan:2017jyc, Benini:2017aed,
 	Aitken:2017nfd, Jensen:2017bjo, Chattopadhyay:2018wkp,
 	Turiaci:2018nua, Choudhury:2018iwf, Karch:2018mer, Aharony:2018npf,
 	Yacoby:2018yvy, Aitken:2018cvh, Aharony:2018pjn, Dey:2018ykx,
 	Chattopadhyay:2019lpr, Dey:2019ihe, Halder:2019foo, Aharony:2019mbc,
 	Li:2019twz, Jain:2019fja, Inbasekar:2019wdw, Inbasekar:2019azv,
 	Jensen:2019mga, Kalloor:2019xjb, Ghosh:2019sqf, Inbasekar:2020hla, Jain:2020rmw, Minwalla:2020ysu, Jain:2020puw, Mishra:2020wos, Minwalla:2022sef, Mehta:2022lgq, Gabai:2022vri, Gabai:2022snc, Gabai:2022mya, Minwalla:2023esg, Jain:2024bza}.
  The 3d bosonization duality is a strong-weak field theory duality, relating the strong coupling limit of one theory to the weak coupling limit of another. Most of the explicit computations aimed at extensively testing 3d bosonization rely on approximation techniques from large-$N$ field theories. While a first-principles proof remains challenging, there is now significant evidence supporting the conjectured duality between $U(N_F)$ (or $SU(N_F)$) Chern-Simons gauge theory coupled to fundamental regular fermions (RF) and $SU(N_B)$ (or $U(N_B)$) Chern-Simons gauge theory coupled to fundamental Wilson-Fisher or critical bosons (CB), along with a specific duality map that relates the mass parameters of the matter fields and the levels and ranks of the respective gauge groups \cite{Aharony:2012ns, Jain:2013py, Jain:2014nza, Choudhury:2018iwf, Minwalla:2020ysu} \footnote{This pair of dual theories, which are also called quasi-fermionic theories, will be the focus of the study in this paper. There is also another more general pair of dual theories known as the quasi-bosonic theories \cite{Minwalla:2020ysu}, which we will not discuss here but can also be studied in the same way.}. 
 	
  \vspace{0.15cm}

 The duality, in the case of massless bosons and fermions, asserts the equivalence between two distinct families of conformal field theories (CFTs), each defined by the rank and level of their respective gauge groups. This duality, however, is also conjectured to remain valid when both families of CFTs are deformed by the introduction of a mass. 
 The exact characteristics of these massive phases, however, depends upon the signs of the mass deformation. The elementary, one-particle, massive, excitations $|\phi^i\rangle$, in a bosonic theory deformed by a positive mass ($m_B^{\text{cri}}>0$), arise from the action of the scalar field $\phi^i$ on the vacuum. 
The label $i$, here, represents a fundamental index related to the associated Chern-Simons gauge group, indicating that these classically spin-zero excitations transform in the fundamental representation of the gauge group. The fermionic dual of this theory is a gauged `free fermion' theory, where the sign of the fermion mass $m_F$ is chosen to match the sign of the fermonic Chern-Simons level $k_F$ \footnote{For a brief summary of the notations and conventions, see the appendix \ref{appA}. For more details about the notations and conventions, refer to \cite{Minwalla:2020ysu, Minwalla:2022sef}.}. In this phase, the elementary, one-particle, massive excitations $|\psi^i\rangle$ are created by the action of the elementary fermionic fields $\psi^i$ on the vacuum. 
These fermionic excitations transform in the fundamental representation of the corresponding Chern-Simons gauge group, and they have the classical spin $\frac{1}{2}\sgn(m_F)$. These fermionic excitations $|\psi^i\rangle$, in a way\footnote{For instance, under the conjectured duality map, the `physical' quantities (i.e., the spacetime quantum numbers) - such as the exact pole masses and renormalized spins - match at the level of the single-particle states. Additonally, the $S$-matrices corresponding to two-particle scattering states are known to map on both sides of the duality. Furthermore, the thermodynamical ensembles of both bosons and fermons manage to exhibit the same physics.}, correspond under a conjectured duality map to the massive scalar excitations $|\phi^i\rangle$
discussed above
\cite{Aharony:2012ns, Jain:2013py, Jain:2014nza, Choudhury:2018iwf, Minwalla:2020ysu}. 

\vspace{0.15cm}

In contrast, when the bosonic theory is deformed by a negative mass ($m_B^{\text{cri}}<0$), it results in a type of `gauged ferromagnetic' phase \cite{Aharony:2012nh,Choudhury:2018iwf}. In this phase, the `condensation of spins' initiates the Higgs mechanism, which `absorbs' the `ferromagnetic' Goldstone bosons. This process classically transforms their degrees of freedom into a set of massive fields, as the Goldstone bosons' degrees of freedom are integrated into the dynamics of the massive fields.\footnote{As a consequence, the original symmetry associated with the Goldstone modes is broken, leading to a more stable ground state characterized by these massive fields.} This set of massive fields features a multiplet of $W_\mu^i$ bosons\footnote{The little group spin of the on-shell $W_\mu$ bosons equals $\sgn(\kappa_B)$, where, $\kappa_B$ denotes the `renormalized' level for the bosonic Chern-Simons gauge fields, as defined in \eqref{renlevels}. $\kappa_B$ appears in the Lagrangian \eqref{hpactioncl} for the critical bosons in the higgsed phase.} which transform under the fundamental representation of the unbroken portion of the gauge group, alongside a single, neutral, massive $Z_\mu$ bosons that remains uncharged w.r.t. this unbroken subgroup. The dual description of this phase is given by the gauged free fermions, where the sign of the mass of the fermion $m_F$ in this case is chosen to match the opposite of the sign of the Chern-Simons level $k_F$. Once again, the fundamental massive excitations are generated by the elementary fermionic operators, however the classical spin of these excitations are now given by $-\frac{1}{2}\text{sgn}(m_F)$. These excitations transform under the fundamental representation of the Chern-Simons gauge group coupled to fermions, and under the duality, in a sense maps to the $W_\mu^i$ bosons.  Apparently, there seems to be no fermionic analogue of the massive $Z_\mu$ boson to which it can be mapped under Bose-Fermi duality. 

\vspace{0.15cm} 

The Bose-Fermi duality implies that the states in the bosonic theory have a correspondence with those in the fermionic theory. 
Although these correspondences are gratifying, they raise an important question: having identified the bosonic duals of all the obvious fermionic excitations, what is the fermionic dual of the massive, vector Z boson? Furthermore, what are the characteristics of the $Z$ boson, even within the bosonic theory itself? This is the question we will explore and resolve in this paper. 
The paper is structured as follows: In Section \ref{sec2}, we lay the groundwork for the computations needed to explore aspects of the $Z$-boson. Section \ref{sec3} explores the analytic structures of the structure functions for even and odd components of the two-point function of the current. In Section \ref{sec4}, we analyze the spectral decomposition of these components in the Kallen-Lehmann type representation. In Section \ref{dualizedsmatrixwwww}, we provide a brief study of the dualized $2\to 2$ $S$-matrices in the singlet channel in the Higgsed phase. 
Finally, Section \ref{sec5} contains some concluding remarks.

\section{Setup of the computation}\label{sec2}


In the theory of critical bosons transforming under the fundamental representation of the $SU(N_B)_{k_B}$ Chern-Simons gauge group, two distinct phases can occur\footnote{Two distinct phases correspond to two different possible choices of the sign of the mass parameter $m_B^{\text{cri}}$ appearing in the Lagrangian \eqref{cbaction} corresponding to the critical bosons coupled to Chern-Simons gauge fields.}: (i) a phase with complex scalars that remains un-Higgsed, and (ii) a Higgsed phase featuring $W_\mu$ and $Z_\mu$ bosons coupled to the unbroken $SU(N_B-1)_{k_B}$ Chern-Simons gauge group \cite{Aharony:2012nh, Choudhury:2018iwf}. Under Bose-Fermi duality, the theory of the critical bosons map to the thoery of regular fermions coupled to $U(N_F)_{k_F}$ Chern-Simons gauge group \cite{Aharony:2012ns, Jain:2013py, Jain:2014nza, Minwalla:2020ysu}. The mass parameter of the regular fermions appearing in the Lagrangian \footnote{The classical Lagrangian for the regular fermions coupled to Chern-Simons gauge fields is given by \eqref{rfaction}, with the mass parameter denoted by $m_F$. } can have two possible choices of the signs. When the sign of the fermion mass matches the sign of the Chern-Simons level, the 
fermionic one-particle states correspond under duality to the 
complex scalars in the un-Higgsed phase. On the contrary, when the sign of the mass of the fermions is chosen to be opposite of the sign of the Chern-Simons level coupled to fermions, the 
fermionic one-particle states map to the vector $W$ bosons. 

\vspace{0.15cm} 

At this point, one might wonder whether there exists a corresponding state on the fermionic side to which the massive $Z$ boson, in the Higgsed phases of the bosons, can be mapped. In other words, is the $Z$ boson an elementary, stable particle? \footnote{In the Standard Model in $(3+1)$ dimensional spacetime, the $W$-boson and $Z$-boson are fundamental particles and not composite, although the $W$ boson and $Z$ boson decay because of their large masses into other lighter particles due to the available allowed interactions in the Standard Model.} Or, since the vector $Z$ boson is neutral under the unbroken subgroup, does it correspond under duality to a composite fermion-antifermion state, or some other gauge invariant, composite state?

\vspace{0.15cm} 

A significant insight into this question emerges from the fact that, within the classical theory with an action \eqref{hpactioncl}, the mass of the $Z$ boson, denoted as $m_{Z}$, is found to be twice that of the $W$ boson, represented as $m_{W}$ \cite{Choudhury:2018iwf}. In other words, 
\begin{equation}\label{clmassrels}
	m_{Z}= 2 m_{W} \ . 
\end{equation}
Since, the massive $Z$ boson is a singlet under the unbroken subgroup of the bosonic Chern-Simons gauge group, this mass relation suggests that it can plausibly be regarded classically as the $W\overline{W}$ two-particle `marginally' bound state at threshold. Within the framework of duality, the $W$ bosons map to the fundamental fermions. It implies that the $Z$ boson could possibly be interpreted as dual to a bound state created by a fermion and an anti-fermion at threshold energy
\footnote{One might question the apparent contradiction between the anticipation of the appearance of a `marginally' bound state at threshold in the fermionic theory and the conclusion in \cite{Frishman:2014cma}, which states that large $N$ Chern-Simons matter theories with massive fundamental fermions (in regular fermionic theory) do not support bound states. However, at first glance, this does not seem to conflict with \cite{Frishman:2014cma}'s conclusion.
	
	To clarify, a bound state at threshold is not considered a `genuine' bound state, despite sharing some similarities. A genuine bound state (e.g., a hydrogen atom) involves particles bound by an interaction with non-zero binding energy, remaining bound in the absence of external influences.
	
	In contrast, a bound state at threshold occurs when the binding energy approaches zero, meaning the system's total energy equals the energy it would have if the particles were free and non-interacting. This state is marginally bound and unstable, as any small perturbation could cause the particles to become unbound. It is more accurately described as a resonance than a true bound state. 
	
	We will leave a more detailed analysis of whether this is indeed the correct dual interpretation for future work, which may also require input from finite $N$ corrections. 
	
	} \footnote{Since there is apparently no classical fermion analog of the $Z$ boson, how could this be reconciled with duality? Several possibilities arise: 
	First, the fermionic theory may develop a bound state, similar to positronium, at each finite value of the fermionic coupling $k_F$ and $N_F$, with the binding energy approaching zero as $k_F\to \infty$. In this case, the bound state would correspond to the dual $Z_\mu$-boson, which would be stable at all values of $N_B$ and $k_B$, disappearing only in the fermionic zero coupling limit, $\lambda_F\to 0$. 
	
	Second, while the $Z_\mu$-boson is stable in the limit $\lambda_B\to 0$, it might become unstable at any finite value of $k_B$, regardless of how small. In this scenario, the $Z$ boson never exists as a stable particle, at any finite value of $N_B$ and $k_B$ explaining the absence of a dual excitation of fermionic side. 
	
	Another possibility involves a trade-off between the first two scenarios. The $Z_\mu$ boson might turn out to be stable at small values of $\lambda_B$, but unstable in the complementary range of parameters corresponding to small values of $\lambda_F$.  
	
	While our large $N$ analysis in this paper aligns with the second possibility, providing a more precise answer likely requires an analysis at finite $N$, which we will leave for future work. \label{finitenfn}}.


\vspace{0.15cm} 

Given that the $Z$ boson at the classical level exists as a bound state at threshold, its behavior upon the introduction of bosonic gauge coupling remains uncertain. The effects of this coupling could either stabilize the threshold bound state into a true bound state or destabilize it, resulting in an unstable resonance. Determining which outcome occurs is a complex dynamical issue that cannot be addressed through general principles alone.

\vspace{0.15cm} 

In this paper, we investigate this question within the framework of the strict 't Hooft large $N$ limit, with calculations valid at all orders in the 't Hooft coupling to ascertain the fate of the dynamical $Z$ boson pole. As we will elaborate further, our findings indicate that the $Z$ boson indeed transforms into an unstable resonance, but one with unique characteristics. 
Before delving into this phenomenon in detail, we first describe the specific quantity we study to characterize the $Z$ boson.

\vspace{0.15cm} 

We consider the simplest possible dual pair of non-supersymmetric, massive Chern-Simons matter theories, namely the critical bosonic theories and the regular fermionic theories, with (special) unitary gauge groups. In the strict large $N$ limit, the classical (Euclidean) Lagrangian for the critical bosonic theory with $SU(N_B)$ Chern-Simons gauge fields is given by \eqref{cbaction}, whereas for the critical fermionic theory with $U(N_F)$ Chern-Simons gauge fields, the corresponding classical (Euclidean) action is given by \eqref{rfaction}. 

\subsection{Observables of interest}

Both the dual pair comprising critical boson theories and regular fermion theories exhibit invariance under a global $U(1)$ symmetry. 
In the context of regular fermionic matter theory coupled to Chern-Simons gauge fields, the associated conserved current can be written as follows
\begin{equation}\label{u1fcxspace}
	J_\mu^F =\i \bar{\psi} \gamma_\mu \psi \ .  
\end{equation}
On the other hand, in the unhiggsed phase of scalar bosonic theory, the corresponding conserved current can be expressed as\footnote{The covariant derivative operator appearing in \eqref{u1scalarcurrent} is defined as 
	$D_\mu \equiv \partial_\mu -\i A_\mu \ ,   $
	where, $A_\mu$ is the corresponding Chern-Simons gauge field.}
\begin{equation}\label{u1scalarcurrent}
	J_\mu^B = \i \Big(\big(D_\mu \bar{\phi}\big)\phi - \bar{\phi} \big(D_\mu\phi \big) \Big) \ . 
\end{equation}
 Under Bose-Fermi duality, the two currents $J_\mu^F$ and $J_\mu^B$ are conjectured to be dual to each other. 
 In the higgsed phase the Higgs mechanism causes the scalar fields to acquire a vacuum expectation value (vev). When expressed in the unitary gauge, this expectation value can be represented in the following manner \cite{Choudhury:2018iwf}:
\begin{equation}\label{unitarygaugevev}
	\phi^i \equiv \delta^{i N_B} v\sqrt{|\kappa_B|} = \delta^{i N_B} \sqrt{\frac{N_B}{4\pi}|m_B^\text{cri}|} \ .  
\end{equation}
Here, $v$ is associated with the vev of the scalar field, and is related to the magnitude of the mass parameter $m_B^{\text{cri}}$ through the second equality in \eqref{unitarygaugevev}. It is noteworthy that this vev is intricately linked to the mass of the vector $W$ boson which in turn implies that this vev also influences the vector $Z$ boson. The precise relationship between the large $N$ exact mass of the $W$ boson and the vev is given by the following formula \cite{Choudhury:2018iwf} \footnote{For later convenience, for a generic allowed value of $\lambda_B$, the mass of the $Z$ boson is chosen to be defined to be as $2c_B$. As we will see this is the peak position of the resonance. This definition appropriately simplifies to the tree level pole mass of the $Z$ boson as the 't Hooft coupling parameter approaches to zero.}:
\begin{equation}\label{cBvevlambda} 
	c_B =\frac{2\pi v^2}{1-|\lambda_B|/2} \ . 
\end{equation}


\vspace{0.15cm} 

It follows from \eqref{u1scalarcurrent} and \eqref{unitarygaugevev} that in the higgsed phase, the current for the scalars, i.e., $J^B_\mu$ reduces to 
\begin{equation}\label{u1currenthiggsed}
	J_\mu^H = -2|\kappa_B| v^2 ~ Z_\mu \ . 
\end{equation}
In this context, the color singlet, massive, vector $Z$ boson field, denoted by $Z_\mu$, corresponds to the `$N_BN_B$' color component of gauge field \footnote{Before Higgsing, the gauge field transforming in the adjoint representation of $SU(N_B)$ gauge group is the $N_B\times N_B$ matrix valued field.}. It is important to note that the $U(1)$ current operator turns out to be proportional to $Z$ boson. We follow the normalizaiton of the fields as in \cite{Choudhury:2018iwf}, and the $Z$ boson field is normalized such that the tree-level two-point function takes the following form \cite{Choudhury:2018iwf}: 
\begin{equation}\label{2pttreez}
	\big\langle Z_\alpha(q) Z_\beta(p) \big\rangle =(2\pi)^3 \delta^{(3)}(q+p)~  \frac{2\pi m_Z}{|\kappa_B|(q^2+m_Z^2)} ~  \bigg( \delta_{\alpha\beta} - \sgn(\kappa_B)\epsilon_{\alpha\beta\mu}\frac{q^\mu}{m_Z} +\frac{q_\alpha q_\beta}{m_Z^2}\bigg) \ . 
\end{equation}
This implies that the two-point function of the current \eqref{u1currenthiggsed} will contain an additional overall factor of $|\kappa_B|$ in the numerator. However, we consider an overall normalization factor (including $|\kappa_B|$ along with other numerical factors for later convenience) as below in \eqref{normjz} to help us take a smooth zero coupling limit $\lambda_B\to 0$ \footnote{The 't Hooft coupling parameter is defined by $\lambda_B =\frac{N_B}{\kappa_B}$. Hence, $\lambda_B\to 0$ corresponds to $\kappa_B\to \infty$ while keeping $N_B$ finite.}, while also removes an overall factor of $\kappa_B$ from the tree level result. To put it another way, the two-point correlation function of the normalized current 
\begin{equation}\label{normjz}
	\tilde{J}_\alpha = \sqrt{\frac{2\pi}{|\kappa_B|}} ~ J_\alpha^H  \ , 
\end{equation}
as the 't Hooft limit approaches to zero, must reduce to the following normalized tree level two-point correlation function: 
\begin{equation}\label{2ptnormtreej}
	\big\langle \tilde{J}_\alpha(q) \tilde{J}_\beta(p) \big\rangle =(2\pi)^3 \delta^{(3)}(q+p)~  \frac{m_Z^3}{q^2+m_Z^2} ~  \bigg( \delta_{\alpha\beta}\Big(1+\frac{q^2}{m_Z^2}\Big) - \sgn(\kappa_B)\epsilon_{\alpha\beta\mu}\frac{q^\mu}{m_Z} +\frac{q_\alpha q_\beta-q^2\delta_{\alpha \beta}}{m_Z^2}\bigg) \ . 
\end{equation}

\vspace{0.15cm}

An effective method of exploring the behaviour of the $Z$ boson beyond the classical limit is to compute the two-point function of the normalized current operator $\tilde{J}_\alpha$ at all orders in the 't Hooft coupling in the exact large-$N$ limit. This approach allows us to systematically investigate how the classical pole evolves as we move into the quantum regime.

\vspace{0.15cm} 

To evaluate the two-point current correlation function of the normalized current operator \eqref{normjz}, a direct approach would involve starting with the classical Lagrangian of the bosonic theory in the higgsed phase \eqref{hpactioncl}, and then compute by summing all the planar diagrams to all orders in the 't Hooft coupling limit $\lambda_B$. Although this computation is certainly possible, it is not trivial, and it comes with its own set of complexities. Rather than grappling with intricacies of the computation, we opt for a more efficient strategy in this paper. 

\vspace{0.15cm} 

 We leverage the fact that the two-point function corresponding to the fermionic current operator $\i \bar{\psi}_i \gamma_\alpha \psi^i$, which is dual to the current $J_\mu^H$ (which we rescale as $\tilde{J}_\alpha$ in \eqref{normjz} for a smooth zero coupling limit), has already been computed in the works of \cite{Gur-Ari:2016xff, Mishra:2020wos}, which we have summarized in Appendix \ref{apprf}. This allows us to circumvent some of the more challenging aspects of the direct calculation, while still obtaining the necessary insights into the behaviour of the $Z$ boson. 
 
 \vspace{0.15cm}
 
 Assuming the Bose-Fermi duality holds true, we can directly apply the findings of \cite{Gur-Ari:2016xff, Mishra:2020wos} along with the standard duality map to extract for the prediction of the large-$N$ exact two-point correlation function of the current operator outlined in \eqref{normjz}. This approach allows us to efficiently extract the relevant results, leading to the following (Euclidean) expression\footnote{In addition to the terms appearing in \eqref{2ptpred}, the exact expression of the current two-point function in this case also contains an identity piece, i.e., proportional to $\delta_{\alpha\beta}$, independent of the momentum. This extra term originates from the coupling of bosons with the background gauge field corresponding to the global $U(1)$ symmetry. We have not included in \eqref{2ptpred} but is required for the matching with the classical propagator of the $Z$ boson at the bosonic weak coupling limit, and hence is included in \eqref{includeddab}. However, since this identity term does not contribute to the analytic structure relevant to us, we do not include it here.} 
 \begin{equation}\label{2ptpred} 
 	\Big\langle \tilde{J}_\alpha(q) \tilde{J}_\beta(p) \Big\rangle  = (2\pi)^3 \delta^{(3)}(q+p) ~ \bigg(\tau_Z(p) ~(-p_\alpha p_\beta +p^2\delta_{\alpha\beta}) - \kappa_Z(p)~\sgn(\kappa_B) \epsilon_{\alpha\beta\mu}p^\mu \bigg) \ , 
 \end{equation}
 where, the functions $\tau_Z$ and $\kappa_Z$ are the coefficient structure functions corresponding to the even and the odd tensor structures, respectively. Although it may seem a misnomer, we refer to these functions as the even and odd component structure functions (or sometimes simply, even and odd components, in short) throughout this paper. The explicit forms of these two functions are  as follows \footnote{From the perspective of the CFTs prior to the mass deformations, one might discard the parity-odd contributions as they correspond to contact terms. However, for completeness, we retain the contributions from the odd components here.}
 \begin{eqnarray}\label{pepohiggsztwopt}
 \begin{split} 
 	\tau_Z(p)  = ~ &  \frac{\i}{16p^3} \bigg[\frac{(2c_B-\i p)^4}{(p^2+4c_B^2)}\Big(\frac{2c_B+\i p}{2c_B-ip}\Big)^{|\lambda_B|}-\frac{(2c_B+\i p)^4}{(p^2+4c_B^2)}\Big(\frac{2c_B-\i p}{2c_B+ip}\Big)^{|\lambda_B|}\bigg] \\
 	 & - \Big(1-\frac{|\lambda_B|}{2}\Big)\frac{c_B}{p^2} \  ,  \\ 
 	\kappa_Z(p)  = ~ &  \frac{1}{16p^2} \bigg[\frac{(2c_B-\i p)^4}{(p^2+4c_B^2)}\Big(\frac{2c_B+\i p}{2c_B-\i p}\Big)^{|\lambda_B|}+\frac{(2c_B+\i p)^4}{(p^2+4c_B^2)}\Big(\frac{2c_B-\i p}{2c_B+\i p}\Big)^{|\lambda_B|}\bigg] \\ 
 	&  - \frac{1}{8} \Big(1+\frac{4c_B^2}{p^2}\Big) + \frac{1}{4} \ . 
 \end{split} 
 \end{eqnarray}
 In the above two expressions \eqref{pepohiggsztwopt}, $p$ appearing in the arguments of the functional forms on the LHS as well as in the explicit expressions on the RHS is the modulus of the Euclidean 3-momentum vector $p_\mu \equiv (p_1, p_2, p_3)$, i.e., 
 \begin{equation}\label{eqmommod}
 	p=+\sqrt{p^2} = +\sqrt{p_1^2+p_2^2+p_3^2} \ , 
 \end{equation}
 To verify our findings, we observe that as the 't Hooft coupling approaches zero, i.e., $|\lambda_B|\to 0$, the even and odd components simplify in the following manner
 \begin{equation}\label{zerocouplinglimitpepo}
 	\tau_Z(p) = -\frac{m_Z}{(p^2+m_Z^2)} \ ,  ~ ~ ~ \kappa_Z(p) = -\frac{m_Z^2}{(p^2+m_Z^2)} + \frac{1}{4} \ . 
 \end{equation}
It is important to highlight that the only analytic structure present is the pole located at $p^2=-m_Z^2$. This behaviour is consistent with what we expect from the classical propagator of the $Z$ boson. 

\vspace{0.15cm} 

Before proceeding, it's worth noting the discrepancies among the terms. Up to constant terms proportional to $\delta_{\alpha\beta}$ and polynomial terms in $p^\mu$ the expression in \eqref{zerocouplinglimitpepo} aligns with the normalized propagator of the $Z$-boson derived from the classical Lagrangian, as shown in \eqref{2ptnormtreej}. The specific disagreement at the tree level can be summarized as follows
\begin{equation}\label{mismatchtreelevel}
	 (2\pi)^3 \delta^{(3)}(q+p) ~ \bigg(m_Z \delta_{\alpha \beta} +\frac{\sgn(\kappa_B)}{4}\epsilon_{\alpha\beta\mu}p^\mu  \bigg) \ .
\end{equation}
Nonetheless, the discrepancy is not a cause for concern, as the polynomial term in question is a contact term that can be ignored. Furthermore, its presence does not impact the conservation laws of the current. The appearance of the constant $\delta_{\alpha \beta}$ term is due to the fact that the two-point correlation function $\langle J_\alpha^H J_\beta^H\rangle$ that we considered above is not complete answer, since it is not gauge invariant under the background gauge symmetry. The background $U(1)$ gauge invariant quantity on the other hand is \footnote{For a quick but schematic way to obtain this additional second term, see for instance in \cite{Gur-Ari:2016xff} in the eqn. 80, and combine it with the unitary gauge \eqref{unitarygaugevev}. The origin of the second term proportional to $\delta_{\alpha\beta}$ is due to the fact that in the presence of a background gauge field $\mathcal{A}_\mu$ corresponding to the $U(1)$ symmetry, there is an additional term in the Lagrangian of the form $|\phi|^2\mathcal{A}_\alpha\mathcal{A}^\alpha$, which in the Higgsed phase takes the following form $|\kappa_B| v^2 \mathcal{A}_\alpha\mathcal{A}^\alpha$. To compute the two-point function of the $U(1)$ current, one needs to take derivative w.r.t. $\mathcal{A}_\alpha$ twice, which gives rise to the second term. It can also be rederived directly in the higgsed phase considering the corresponding background $U(1)$ gauge symmetry \cite{Choudhury:2018iwf} and the associated two-point function of the current operator.}
\begin{equation}\label{gitwopt}
	\langle J_\alpha^H(x) J_\beta^H(y) \rangle -2|\kappa_B|v^2 \delta_{\alpha\beta}\delta^{(3)}(x-y) \ . 
\end{equation}
If we consider the normalization used in \eqref{normjz} to take into account of the normalization of the constant term as well, the normalized, background gauge-invariant two-point correlator is given by 
\begin{equation}\label{includeddab}
	\mathcal{G}_{\alpha\beta}(x,y)= \langle \tilde{J}_\alpha(x) \tilde{J}_\beta (y) \rangle - m_Z \delta_{\alpha\beta} \delta^{(3)}(x-y)  \ .
\end{equation}
This exactly matches with the constant term appearing in \eqref{mismatchtreelevel} and hence is not a problematic term at all.

\section{Analytic structures of the structure functions of even and odd components}\label{sec3}
\subsection{Analytic continuation}
The results for the structure functions associated with the even and odd components presented in \eqref{pepohiggsztwopt} are expressed in the Euclidean signature. However, for various applications - such as studying the analytic structure of correlation functions and understanding their singularities - it is crucial to obtain the corresponding results in the Lorentzian signature. 

\vspace{0.15cm} 

To transition from the Euclidean to the Lorentzian signature, we perform an analytic continuation represented by the transformation for the third component $p_3$ of the Euclidean 3-momentum as below
\begin{equation}\label{wickrot}
	p_3 \to -\i p^0
\end{equation}
where $p^0$ denotes the zeroth (i.e., the time) component of the momentum in Lorentzian signature. The momentum in the Lorentzian signature is labelled as $p^\mu\equiv (p^0, p_1, p_2)$ \footnote{We adopt the mostly plus signature $\eta_{\mu\nu}\equiv \text{diag}(-1, +1, +1)$ for the Minkowski background spacetime. The Euclidean background metric is  $\delta_{\mu\nu}\equiv \text{diag}(+1, +1, +1)$.}. In particular the square of the 3-momentum, defined in the Euclidean signature in \eqref{eqmommod}, in the Lorentzian signature, takes the following form 
\begin{equation}\label{euctolormomsq}
	p^2 = p_1^2+p_2^2+p_3^2 \to   - (p^0)^2 + p_1^2+ p_2^2 \ . 
\end{equation}
Additionally, for convenience, we define the $s$ `Mandelstam' variable as \footnote{To interpret the variable $s$, note that for an on-shell particle with a momentum $p^\mu$, its energy is $E_{\vec{p}}\equiv p^0=\sqrt{m^2+\vec{p}^2}$, where, $\vec{p}\equiv (p_1, p_2)$. This implies $p^2=-m^2$. Therefore accroding to the definition \eqref{defsmand}, $\sqrt{s}$ corresponds to mass.}
\begin{equation}\label{defsmand}
	s \equiv - p^2  \geq 0 \ .
\end{equation}
It is beneficial to re-express the expressions in \eqref{pepohiggsztwopt} in terms of the variable $s$, as it allows us to explore the analyticity structures of these functions, as discussed in the following section. 
Consequently, we derive the results for sttructure functions associated with the even and odd components in the $s$ variable as follows 
\begin{eqnarray}\label{peposplane}
	\begin{split} 
		\tau_Z(s) & = \frac{1}{16s^{3/2}} \bigg[ \frac{(4c_B^2-s)^{3-|\lambda_B|}}{(2c_B+\sqrt{s})^{4-2|\lambda_B|}}-\frac{(2c_B+\sqrt{s})^{4-2|\lambda_B|}}{(4c_B^2-s)^{1-|\lambda_B|}}\bigg]  + \Big(1-\frac{|\lambda_B|}{2}\Big)\frac{c_B}{s} \  ,  \\ \\
		\kappa_Z(s) & =  - \frac{1}{16s} \bigg[\frac{(4c_B^2-s)^{3-|\lambda_B|}}{(2c_B+\sqrt{s})^{4-2|\lambda_B|}}+\frac{(2c_B+\sqrt{s})^{4-2|\lambda_B|}}{(4c_B^2-s)^{1-|\lambda_B|}}\bigg]   - \frac{1}{8} \Big(1-\frac{4c_B^2}{s}\Big) + \frac{1}{4} \ . 
	\end{split} 
\end{eqnarray}
Before proceeding further, we pause to clarify the notations used here. We have denoted the two functional forms $\tau_Z(p)$ in \eqref{pepohiggsztwopt} and $\tau_Z(s)$ \eqref{peposplane} (similarly for $\kappa_Z(p)$ and $\kappa_Z(s)$) with the same function symbol to avoid the notational complexity. However, it should be understood that these represent two distinct functions.

\vspace{0.2cm} 

\subsection{Discontinuities of $\tau_Z$ and $\kappa_Z$ in the complex $s$-plane}
While studying the behavior of the functions in \eqref{peposplane} for real, and positive $s$ is of interest, to explore the analyticity properties of the even and odd components in \eqref{peposplane}, it is beneficial to analytically continue them across the complex $s$-plane by extending the variable $s$ into the complex domain. 
Analytically continuing the components in \eqref{peposplane} across the complex $s$-plane, it is easy to observe that both these functions have branch cuts along the real $s$-axis, starting at $s=4c_B^2$ and extending to infinity for $s>4c_B^2$. This enables us to determine the discontinuities across the branch cut along the real $s$-axis, which in turn allows us to compute the imaginary parts of these components. 

\vspace{0.15cm}

To determine the discontinuities, we use the fact that for a complex function of the form $z^\alpha$ (with non-integer $\alpha$)  having a branch cut along the $-ve$ $\text{Re}[z]$ axis, the discontinuity associated with it across the branch cut is given by 
\begin{equation} 
\text{Disc}(z^\alpha) = 2\i |z|^\alpha \sin(\pi \alpha) \ . 
\end{equation} 
The discontinuity is given by the difference in values as $z$ approaches from above or below the branch cut.
In the complex $s$-plane the discontinuities, across the branch cut along the real axis for $s>4c_B^2$, associated with the even and odd components in \eqref{peposplane} of the two-point current correlation funciton in the higgsed phase, are given by 
\begin{eqnarray}\label{disconpeposplane}
	\begin{split} 
		\text{Disc}(\tau_Z(s)) & = \frac{2\i}{16s^{3/2}} \sin(\pi|\lambda_B|)~ \bigg[ \frac{(s-4c_B^2)^{3-|\lambda_B|}}{(2c_B+\sqrt{s})^{4-2|\lambda_B|}}+\frac{(2c_B+\sqrt{s})^{4-2|\lambda_B|}}{(s-4c_B^2)^{1-|\lambda_B|}}\bigg] \Theta(s-4c_B^2) \  ,  \\ \\
		\text{Disc}(\kappa_Z(s)) & = \frac{2\i}{16s}\sin(\pi |\lambda_B|)~ \bigg[-\frac{(s-4c_B^2)^{3-|\lambda_B|}}{(2c_B+\sqrt{s})^{4-2|\lambda_B|}}+\frac{(2c_B+\sqrt{s})^{4-2|\lambda_B|}}{(s-4c_B^2)^{1-|\lambda_B|}}\bigg]  \Theta(s-4c_B^2)  \ . 
	\end{split} 
\end{eqnarray}
Here, in \eqref{disconpeposplane}, $\Theta(x)$ is the Heaviside unit step function, and its presence ensures that the discontinuities are non-zero only for $s>4c_B^2$ and vanish when $s<4c_B^2$. 

\subsection{Imaginary parts of $\tau_Z$ and $\kappa_Z$ in the complex $s$-plane}
We use the fact that the discontinuity and the imaginary parts of a complex function $f(s)$ in the complex $s$-plane across a branch cut along the real $s$-axis, are related to each other as follows \footnote{For a brief review of the proof of \eqref{defimpart} see the appendix \ref{appdisim}.}
\begin{equation}\label{defimpart}
	\text{Im}(f(s)) = \frac{1}{2\i} ~ \text{Disc}(f(s)) \ . 
\end{equation}
Our aim in the next section will be to determine a Kallen-Lehmann type spectral representation for the large-$N$ exact two-point function. It can be shown that the spectral density function is related to the imaginary part of the corresponding analytically continued coefficient function presented in \eqref{2ptpred}. Typically, it is easier to compute the discontinuity than the imaginary part directly.  Putting together the relationship \eqref{defimpart} and the expressions of the discontinuities presented in \eqref{disconpeposplane}, we extract the corresponding imaginary parts of the structure functions for even and odd components of the two-point correlation function in the complex $s$-plane, which are given by 
\begin{equation}\label{impartpepo}
	\begin{split} 
		\text{Im}(\tau_Z(s)) & = \frac{1}{16s^{3/2}} \sin(\pi|\lambda_B|)~ \bigg[ \frac{(s-4c_B^2)^{3-|\lambda_B|}}{(2c_B+\sqrt{s})^{4-2|\lambda_B|}}+\frac{(2c_B+\sqrt{s})^{4-2|\lambda_B|}}{(s-4c_B^2)^{1-|\lambda_B|}}\bigg] \Theta(s-4c_B^2) \  ,  \\ \\
		\text{Im}(\kappa_Z(s)) & = \frac{1}{16s}\sin(\pi |\lambda_B|)~ \bigg[-\frac{(s-4c_B^2)^{3-|\lambda_B|}}{(2c_B+\sqrt{s})^{4-2|\lambda_B|}}+\frac{(2c_B+\sqrt{s})^{4-2|\lambda_B|}}{(s-4c_B^2)^{1-|\lambda_B|}}\bigg]  \Theta(s-4c_B^2)  \ . 
	\end{split} 
\end{equation}

For analyzing the singular properties of these discontinuities (or rather the imaginary parts), it is useful to define at this stage a dimensionless variable $x=\frac{s}{4c_B^2}-1$ and since the discontinuities are zero for $s<4c_B^2$, we consider the effective range of the dimensionless variable $x$ to be $0<x<\infty$. Since the allowed values of the `t Hooft coupling parameter $|\lambda_B|$ range from $0\leq |\lambda_B|\leq 1$, it is important to note that the discontinuities presented in \eqref{disconpeposplane} or the imaginary parts extracted in \eqref{impartpepo} exhibit a singularity structure of the form $x^{|\lambda_B|-1}$ near $s=4c_B^2$. This singularity can be understood in terms of a distribution, schematically as follows (the exact functional forms of the spectral functions are derived in \eqref{rotrok})
\begin{equation}\label{blowupatthreshold}
 \rho(x)\sim \frac{1}{x^{1-|\lambda_B|}} \ ,
\end{equation} 
which resembles a $\delta$-function singularity, which can also be viewed as a distribution. However, it is evident that this distribution is not normalizable over the range $0<x<\infty$. Nevertheless, to make sense of this, an upper cutoff $x_\text{max}=\Lambda$ can be introduced, where the distribution decays sufficiently so that it can be neglected for the remaining range $\Lambda <x<\infty$.

\section{Spectral decomposition}\label{sec4}
At a finite value of the 't Hooft coupling, $|\lambda_B|$, the structure functions for even and odd components can be expanded in the spectral decomposition in terms of the spectral density functions as follows
\begin{equation}\label{spectraldecdef}
	\tau_Z(p) = \int dm^2 ~  \frac{\rho_\tau(m^2)}{p^2+m^2} \ , ~ ~ ~  ~ ~ \kappa_Z(p) = \int dm^2 ~  \frac{\rho_\kappa(m^2)}{p^2+m^2} \ . 
\end{equation}
In \eqref{spectraldecdef}, we have suppressed the $\i \epsilon$'s in the denominator of the integrands for notational convenience. They can easily be restored via the replacement $p^2\to p^2-\i \epsilon$. It can easily be shown that the spectral functions $\rho_\tau$ and $\rho_\kappa$ defined above are related to the imaginary parts of the corresponding functions $\tau_Z$ and $\kappa_Z$, respectively. The precise relationship between the spectral functions and the corresponding imaginary parts is given by 
\begin{equation}\label{specimrel} 
	\rho_\tau(s) =\frac{1}{\pi} ~ \text{Im}\big(\tau_Z(s)\big) \ , ~ ~ ~  ~ \rho_\kappa(s) =\frac{1}{\pi} ~ \text{Im}\big(\kappa_Z(s)\big) \ . 
\end{equation}
A proof of this relationship is provided in the Appendix \ref{appproofimrho}. 
It follows from \eqref{specimrel} and \eqref{impartpepo} that the explicit expressions for the spectral density functions $\rho_\tau$ and $\rho_\kappa$ associated with the even and odd components, are given by 
\begin{equation}\label{rotrok}
	\begin{split} 
		\rho_\tau(m^2) & = \frac{1}{16\pi m^{3}} \sin(\pi|\lambda_B|)~ \bigg[ \frac{(m^2-4c_B^2)^{3-|\lambda_B|}}{(2c_B+m)^{4-2|\lambda_B|}}+\frac{(2c_B+m)^{4-2|\lambda_B|}}{(m^2-4c_B^2)^{1-|\lambda_B|}}\bigg] \Theta(m^2-4c_B^2) \  ,  \\ \\
		\rho_\kappa(m^2) & = \frac{1}{16\pi m^2}\sin(\pi |\lambda_B|)~ \bigg[-\frac{(m^2-4c_B^2)^{3-|\lambda_B|}}{(2c_B+m)^{4-2|\lambda_B|}}+\frac{(2c_B+m)^{4-2|\lambda_B|}}{(m^2-4c_B^2)^{1-|\lambda_B|}}\bigg]  \Theta(m^2-4c_B^2)  \ . 
	\end{split} 
\end{equation}
The spectral functions $\rho_\tau(m^2)$ and $\rho_\kappa(m^2)$ as given by \eqref{rotrok} exhibit several intriguing characteristics. We should emphasize that, on general grounds, tachyonic excitations are undesirable in the spectrum. Therefore, these spectral functions should be supported on positive values of $s=m^2$. It is noteworthy, first of all, that the spectral functions given by \eqref{rotrok} vanish when $m^2<4c_B^2$. In other words, for both the spectral functions the following condition is satisfied
\begin{equation}
	\rho(m^2) = \begin{cases}
		0 & \text{if $m^2< 4c_B^2$} \ , \\
		\geq 0 & \text{if $m^2> 4c_B^2$} \ . 
	\end{cases}
\end{equation}
Another important point we should highlight here is the absence of the Dirac $\delta$-function in the spectral functions \eqref{rotrok} at $m^2=4c_B^2$, which indicates that the $Z$-boson is not a stable particle for any non-zero value of $\lambda_B$.  The spectral functions, as shown in \eqref{rotrok}, correspond to two-particle $W\overline{W}$ cuts starting at $m^2=4c_B^2$ \footnote{We should highlight that the Kallen-Lehmann representation for unstable particles, in contrast to stable particles, lacks a one-particle pole on the real axis. }.  

\vspace{0.15cm}

At fixed allowed values of $|\lambda_B|$, both the even and odd component spectral functions blow up when $m\to 2c_B$, due to the presence of the factor $(m^2-4c_B^2)^{|\lambda_B|-1}$. 
On the contrary, at a fixed arbitrary value of $m^2$, as the 't Hooft coupling parameter $|\lambda_B|$ approaches zero, the spectral density functions as given in \eqref{rotrok} vanish due to the presence of an overall factor $\sin(\pi |\lambda_B|)$ in both expressions. 

\vspace{0.15cm} 

A rather interesting regime is when $|\lambda_B|$ is infinitesimally small and also $m^2$ is close to but larger than $4c_B^2$. In this regime, to the leading order in $|\lambda_B|$, the spectral density functions simplify as follows:
\begin{equation}\label{simrotrok}
	\rho_\tau \approx |\lambda_B| m_Z ~ \frac{(2m_W)^{-2|\lambda_B|}}{(m^2-4m_W^2)^{1-|\lambda_B|}} \ , ~ ~ ~ ~  \rho_Z \approx  |\lambda_B| m_Z^2 ~ \frac{(2m_W)^{-2|\lambda_B|}}{(m^2-4m_W^2)^{1-|\lambda_B|}} \ .
\end{equation}
The plots (see appendix \ref{appplots} and in particular Figures \ref{fig:rhotplot} and \ref{fig:rhokplot}) of the spectral density functions show that they have a sharp fall near $m^2=4m_W^2$ for small $|\lambda_B|$. This suggests to determine the spectral width of these density functions. We will determine here the width of the spectral density function of the even component. The same can be said for the odd component as well. We consider a small dimensionless parameter $w$ which characterizes the width of the spectral density function corresponding to the even component. At a small value of $|\lambda_B|$, we find that the `normalized' area under the curve of the spectral density function in the range $4c_B^2< m^2 < (1+w)4c_B^2$
\begin{equation}\label{normaucperhot}
	\frac{1}{m_Z} \int_{m_Z^2}^{(1+w)m_Z^2} \rho_\tau(m^2)~dm^2 = w^{|\lambda_B|} \ . 
\end{equation}
For a generic, positive value of $w$, the normalized area under the curve is given by the RHS of \eqref{normaucperhot}, which reduces to unity in the limit, $|\lambda_B|\to 0$. 
This is a sense in which the spectral function reduces to a delta function singularity at $m^2=4m_W^2$ corresponding to a stable bound state at threshold, in the limit $|\lambda_B|\to 0$. In other words, in the bosonic zero coupling limit, the spectral density function approaches to a delta-function type singularity in the limiting sense, and the corresponding normalized delta-function behaviour can be written the following form:
\begin{equation}
	\rho_\tau(m^2) = 2m_W~\delta(m^2-4m_W^2) \ . 
\end{equation}
When $|\lambda_B|$ is small but non-zero, the spectral density of the two-point function, which would typically be sharply concentrated around a very narrow region near $2m_W$, begins to spread out significantly. What is particularly interesting about this broadening is its unusual nature: rather than taking the typical Lorentzian distribution that one might expect, the spectral density instead shows a power-law behaviour. 

\vspace{0.15cm} 

Let us define the width of this distribution as $m_Z^2$ multiplied by the value of $w$ for which the right-hand side of \eqref{normaucperhot} evaluates to $1/e$. This corresponds to the range within which the $\frac{1}{e}$ fraction of the broadened spectral density is contained. Therefore, we find that\footnote{The precise form of the expression in \eqref{widthexp} is contingent upon the specific definition of the width that we adopt. For instance, if we were to define the width as the range that encompasses a fraction $1/u$ of the spectral density, the expression for the width in \eqref{widthexp} would be modified to $w=u^{-1/|\lambda_B|}$. }
\begin{equation}\label{widthexp}
	w = \exp\Big(-\frac{1}{|\lambda_B|}\Big) \ . 
\end{equation}
It is important to note that in the limit $|\lambda_B|\to 0$, the width becomes non-perturbatively small. This leads to the conclusion that, although $Z_\mu$ is unstable for all values of coupling constant $|\lambda_B|$, its lifetime \footnote{The lifetime is of the order of the inverse of the width, i.e., $\tau\sim w^{-1}$. This implies that the lifetime scales as $\tau \sim \exp\Big(\frac{1}{|\lambda_B|}\Big)$. Therefore, in the limit $|\lambda_B|\to 0$, the lifetime grows without bound, which indicates that in the bosonic zero coupling limit, the $Z$ boson becomes stable.} grows non-perturbatively large as the 't Hooft coupling parameter $|\lambda_B|$ approaches zero.

\vspace{0.15cm} 

We would like to emphasize that integrals of the spectral density functions corresponding to both even and odd components exhibit divergence. Specifically, the integrals 
\begin{equation}\label{phoint}
\int_{0}^{\infty} \rho_\tau(m^2) ~ dm^2 ~ ~ ~ ~  \text{and}~ ~ ~ ~  \int_{0}^{\infty} \rho_\kappa(m^2) ~ dm^2
\end{equation} 
diverge, 
when evaluated over the full range of $m^2$, extending from zero to infinity \footnote{However, it is important to note that the effective lower limit of both integrals in \eqref{phoint} is $4c_B^2$, as the corresponding spectral density functions in \eqref{rotrok} vanish for $0<m^2<4c_B^2$.}. 
This suggests that the spectral density functions cannot be normalized in the conventional sense, and therefore, a regularization procedure is necessary to obtain physically meaningful results. One effective method for regularizing these functions is to introduce an upper cutoff, as outlined in \eqref{normaucperhot}. 

\section{A brief study of the dualized $S$-matrix for $2\to 2$ $W\overline{W}$ scattering}\label{dualizedsmatrixwwww}
To explore whether the $W\overline{W}$ spectral cut - observed in the large $N$ exact spectral functions for the two-point function of the $Z$ boson - appears in other observables,, in this Section, we look at the scattering matrix corresponding to the $W\overline{W}\to W\overline{W}$ scattering process. However, instead of directly computing this $S$-matrix in the higgsed phase, we assume the validity of the 3d boson-fermion correspondence in the non-supersymmetric Chern-Simons matter theories both in the higgsed and un-higgsed phases of bosons. This assumption is supported by extensive computational evidence for this duality (see for instance in \cite{Aharony:2012ns, Jain:2013py, Jain:2014nza, Minwalla:2015sca, Choudhury:2018iwf, Dey:2018ykx, Minwalla:2020ysu, Minwalla:2022sef}).

\vspace{0.15cm}

The explicit computation and verification of duality for $2 \to 2$ scattering matrices were presented in \cite{Jain:2014nza} for the un-Higgsed phases of the bosonic matter theory, where the elementary fields are scalars.
The computations for the scattering matrices of regular fermions in \cite{Jain:2014nza} are valid for both possible values of the parameter $\sgn(m_F\lambda_F)$. When  $$\text{sgn}(m_F\lambda_F)=+1 \ , $$ the corresponding fermionic S-matrices are dual to the $S$-matrices of the bosonic scalars.  This duality between regular fermions and critical bosons in the un-Higgsed scalar phase was explicitly checked in \cite{Jain:2014nza}. However, the explicit demonstration of duality for the case where  $$\sgn(m_F\lambda_F)=-1 \ , $$ 
corresponding to the duality between regular fermions and critical bosons in the Higgsed phase - has not yet been performed. 
This is because explicit results for the scattering matrices of the  $W$-bosons, even for $2 \to 2$ scattering, are still unknown.

\vspace{0.15cm} 

 Starting from the fermionic $S$ matrices reported in \cite{Jain:2014nza}, and utilizing the duality, we find a prediction for the $S$ matrices for the $2\to 2$ scattering process of $W\overline{W}$. This prediction allows us to check the appearance of a $W\overline{W}$ spectral cut in the corresponding $S$-martix in the higgsed phase.

\subsection*{Known results for the singlet channel fermionic $S$-matrix}

To proceed, we first recall the $S$-matrices for a fermion and an anti-fermion in the singlet channel \footnote{
	the singlet channel is particularly relevant here because the $Z$ boson is a color singlet under the $SU(N_B-1)$ unbroken gauge subgroup of the critical boson theory. For the $Z$-boson to appear as an intermediate state with a $W\overline{W}$ spectral cut in the scattering process, we should focus on the singlet channel.} reported (as a conjecture with a modified crossing symmetry factor) in \cite{Jain:2014nza} (see for instance at equation (3.14)  or equation (7.14) in \cite{Jain:2014nza}). This singlet channel $S$-matrix for anyonic scattering is explicitly derived in the lightcone Hamiltonian formalism in \cite{Gabai:2022snc}.  The $S$-matrix in the singlet channel is given by 
\begin{equation}\label{singletfersmatrix}
	\begin{split} 
		S_{singlet}^F(s,t,u, \lambda_F)  = & ~  \cos(\pi \lambda_F) I(p_1,p_2,p_3,p_4) ~ + ~ 4\sin(\pi \lambda_F) E(p_1, p_2, p_3) \sqrt{\frac{s t}{u}}  \\ 
		& + 4\sin(\pi \lambda_F) \sqrt{s} ~ \frac{1+\exp\Big(- 2\i (\lambda_F-\sgn(m_F)) \tan^{-1}\big(\frac{\sqrt{s}}{2c_F}\big)\Big)}{1- \exp\Big(- 2\i (\lambda_F-\sgn(m_F)) \tan^{-1}\big(\frac{\sqrt{s}}{2c_F}\big)\Big)} \ . 
	\end{split} 
\end{equation}
This expression is written up to an overall momentum conserving $\delta$-function, which is omitted for convenience. 
Let us now clarify the quantities appearing in the singlet channel $S$-matrix. The 3-momenta of the initial particles are denoted by $p_1$ and $p_2$, while the corresponding 3-momenta for the final particles are $-p_3$ and $-p_4$. For $2\to 2$ scattering, the Mandelstam variables $s, t, ~ \text{and}~  u$ are defined in the standard way as 
\begin{equation}
	s=-(p_1+p_2)^2 , ~ ~ ~ ~ t = -(p_1+p_3)^2 , ~ ~ ~ ~ u = - (p_1+p_4)^2, ~ ~ ~ ~ s+t+u =4c_F^2 \ . 
\end{equation}
Here, $c_F$ represents the large $N_F$ exact pole mass for the fermions \footnote{The exact pole mass for the fermions is related to the bare mass $m_F$ appearing in the Lagrangian \eqref{rfaction} via the equation \eqref{cfpolemass}.}. As defined before, $\lambda_F$ is the `t Hooft coupling parameter for the fermionic theory. The term $I(p_1,p_2,p_3,p_4)$ represents the identity part of the $S$-matrix, whose explicit form is not important for us here (for the detailed expression, see equation (2.14) of \cite{Jain:2014nza}). The $Z_2$ valued variable $E(p_1, p_2, p_3)$, appearing in \eqref{singletfersmatrix}, is defined as 
\begin{equation}
	E(p_1, p_2, p_3) \equiv \sgn (\epsilon_{\mu\nu\rho}p_1^\mu p_2^\nu p_3^\rho)  = \pm 1 , ~ ~ ~ \text{with} ~ ~ \epsilon_{012} = -\epsilon^{012} =  1 \ .  
\end{equation}
Note that $E(p_1, p_2, p_3)$ measures the handedness of the triad of vectors $p_1, p_2, ~ \text{and}~ p_3$. The appearance of $E(p_1,p_2,p_3)$ in the $2\to 2$ scattering matrix alongside the usual Mandelstam variables $s, t, ~ \text{and} ~ u$ highlights a distinct kinematical feature in $(2+1)$-dimensions. 

\vspace{0.15cm} 

Using of the conversion formula \eqref{exptopowconvform}, and performing an analytic continuation to the physical domain $\sqrt{s}>2c_F$ in the Minkowski space, 
we find that the singlet channel $S$-matrix for the $\psi \bar{\psi}\to \psi\bar{\psi}$ scattering can be rewritten in the following simplified form 
\begin{equation}\label{singletfersmatrixsim}
	\begin{split} 
		S_{singlet}^F(s,t,u, \lambda_F)  = & ~  \cos(\pi \lambda_F) I(p_1,p_2,p_3,p_4) ~ - ~ 4\i \sin(\pi \lambda_F) E(p_1, p_2, p_3) \sqrt{\frac{s t}{u}}  \\ 
		& -4\i \sin(\pi \lambda_F) \sqrt{s} ~ \frac{(2c_F+\sqrt{s})^{\lambda_F-\sgn(m_F)}+(2c_F-\sqrt{s})^{\lambda_F-\sgn(m_F)}}{(2c_F+\sqrt{s})^{\lambda_F-\sgn(m_F)}-(2c_F-\sqrt{s})^{\lambda_F-\sgn(m_F)}}
	\end{split} 
\end{equation}
By applying the Bose-Fermi duality map relevant for the higgsed phase (i.e., with $\sgn(m_F\lambda_F)=-1$), we obtain the corresponding singlet channel $S$-matrix for the $W\overline{W}\to W\overline{W}$ scattering in the higgsed phase which is given by \footnote{Also we absorb an overall minus sign into the definition of the $S$-matrix of the higgsed phase in \eqref{higgsedphasepred}, since the overall minus sign of the $S$-matrix is merely a choice of the convention and is physically irrelevant.}
\begin{equation}\label{higgsedphasepred}
\begin{split}
	S_{singlet}^H(s,t,u, \lambda_B)  =  &   \cos(\pi \lambda_B) I(p_1,p_2,p_3,p_4) ~ - ~  4\i \sin(\pi \lambda_B) E(p_1, p_2, p_3) \sqrt{\frac{s t}{u}}  \\ 
	& + 4\i ~\sgn(\lambda_B) \sin(\pi \lambda_B)\sqrt{s} ~ \frac{(2c_B+\sqrt{s})^{2-|\lambda_B|}+(2c_B-\sqrt{s})^{2-|\lambda_B|}}{(2c_B+\sqrt{s})^{2-|\lambda_B|}-(2c_B-\sqrt{s})^{2-|\lambda_B|}} \ . 
\end{split}
\end{equation}
From the second line of this expression \eqref{higgsedphasepred},  it is evident that there is  a spectral cut starting at $\sqrt{s} =2c_B$ and extending to infinity along the positive real axis. 
This suggests that $Z$ boson may show up in the singlet channel S-matrix of $W\overline{W}\to W\overline{W}$. However, we should highlight that unlike the blow-up behaviour in the spectal density, as $s\to 4c_B^2$ (as pointed out in \eqref{blowupatthreshold}), the $S$-matrix in \eqref{higgsedphasepred} behaves smoothly as $s\to 4c_B^2$.  
In particular, as $s\to 4c_B^2$, the expression \eqref{higgsedphasepred} reduces to 
\begin{equation}\label{higgsedsmatthreshold}
\begin{split}
	S_{singlet}^H(s,t,u, \lambda_B)  =  &   \cos(\pi \lambda_B) I(p_1,p_2,p_3,p_4) - 4\i \sin(\pi \lambda_B) \bigg(E(p_1, p_2, p_3) \sqrt{\frac{s t}{u}}  - \sgn(\lambda_B) \sqrt{s} \bigg) ~ 
\end{split}
\end{equation}

\subsection{An observation near $s\to 4m_W^2$ at very weak coupling}
The appearance of the two-particle $W\overline{W}$ spectral cut in the expression of the singlet channel $S$-matrix \eqref{higgsedphasepred}, might lead one to expect an approximate pole at $s=4m_W^2$ in the limit $|\lambda_B|\to 0$, when $Z$ boson is stable. However, the expression in \eqref{higgsedsmatthreshold} contradicts this expectation, showing that there is no pole at $s=4m_W^2$ in \eqref{higgsedphasepred} in the limit $|\lambda_B|\to 0$. This could seem contradictory, especially since the tree-level $W\overline{W}\to W\overline{W}$ scattering in the singlet channel due to a $Z$ boson exchange, governed by the classical action \eqref{hpactioncl}, would typically generate an $S$-matrix with pole contribution. However, this is misleading, as explained in the footnote \ref{fndwz}. The associated on-shell three-point function $Z\overline{W}W$ in the corresponding `cut diagram' governed by the action \eqref{hpactioncl} vanishes at tree level, thus eliminating a pole contribution. 

\subsection{Comments on the non-relativistic limit and the Schrodinger problem}
As discussed earlier, in the limit $\sqrt{s}\to 2c_B$, the singlet channel $S$-matrix in the Higgsed phase simplifies to the form given in \eqref{higgsedsmatthreshold} with an interesting feature that there is no pole at $s=4c_B^2$. However, there is another noteworthy effect that arises when considering the non-relativistic limit $\sqrt{s}\to 2c_B$.  To explore this, let us focus on the term in the second line of \eqref{singletfersmatrixsim}, which we denote as $\tilde{T}_{singlet}^F$:
\begin{equation}\label{tfsinglet}
\tilde{T}^{F}_{singlet} = -4\i \sin(\pi \lambda_F) \sqrt{s} ~ \frac{(2c_F+\sqrt{s})^{\lambda_F-\sgn(m_F)}+(2c_F-\sqrt{s})^{\lambda_F-\sgn(m_F)}}{(2c_F+\sqrt{s})^{\lambda_F-\sgn(m_F)}-(2c_F-\sqrt{s})^{\lambda_F-\sgn(m_F)}} \ .  \\ 
\end{equation}
When we dualize this part of the singlet channel $S$-matrix to the un-higgsed phase (i.e., setting $\sgn(m_F \lambda_F) = +1$) and take the non-relativistic limit $\sqrt{s} \to 2c_B$, the term reduces to
\begin{equation}
\tilde{T}^{uH}_{singlet} =  8\i c_B \sin(\pi |\lambda_B|) \ . 
\end{equation}
Conversely, when we dualize to the Higgsed phase (with $\sgn(m_F \lambda_F) = -1$) and take the same limit, the term becomes
\begin{equation}
\tilde{T}^{H}_{singlet} =  - 8\i c_B \sin(\pi |\lambda_B|) \ . 
\end{equation}
Therefore, we observe a relative sign change in this part of the $S$-matrix when transitioning from the un-higgsed to the higgsed phase in the non-relativistic limit

\vspace{0.15cm} 

On general grounds, we anticipate that the $2\to 2$ scattering matrix \eqref{higgsedsmatthreshold} which is obtained from \eqref{higgsedphasepred} in the near threshold limit 
\begin{equation}
\frac{\sqrt{s}-2c_B}{2c_B} \to 0 \ , 
\end{equation}
can be reproduced by an appropriate Schrodinger equation.  This Schrodinger equation describes the motion of a non-relativistic particle in the background of a pointlike magnetic flux at the origin of space. The boundary condition near $r=0$ forces the partial wave corresponding to the lowest angular momentum in the $s$-wave sector to take the following form (for more details, see \cite{Dandekar:2014era} and the references therein)
\begin{equation}\label{acbbc}
\psi_0(r) \propto \bigg(r^{|\lambda_B|}+\frac{w R^{2|\lambda_B|}}{r^{|\lambda_B|}} \bigg) \ , 
\end{equation}
where, $R$ is an arbitrary scale and $w$ is a free parameter known as the self-adjoint extension parameter. The boundary condition described in \eqref{acbbc} is referred to as the Amelino-Camelia-Bak (ACB) boundary conditions.  For the Hamiltonian to be self-adjoint, the quantity $wR^{2|\lambda_B|}$ must be specified once and for all. Notably, two special values of $w$, namely $w=0$ and $w=\infty$, correspond to a Schrodinger problem with no scale.

\vspace{0.15cm}

It was demonstrated, in \cite{Dandekar:2014era}, that the non-relativistic problem describing $\bar{\phi}\phi$ scattering in the un-Higgsed phase corresponds to a Schrodinger problem with $w=0$. Specifically, the non-relativistic $S$-matrix for the scattering of two-anyons in the $s$-wave sector, obtained by solving the Schrodinger equation with the boundary condition \eqref{acbbc} is given (in the center of mass frame with scattering angle $\theta$) by (see for instance in equation (1.15) of \cite{Dandekar:2014era})
\begin{equation}\label{tnr}
T_{NR} =  -16\pi \i c_B (\cos(\pi \lambda_B)-1) \delta(\theta) + 8\pi \i c_B \sin(\pi \lambda_B) \text{Pv}\big(\cot\frac{\theta}{2} \big) +T_{w} \ , 
\end{equation}
where the term $T_w$ in the non-relativistic $S$-matrix contains the self-adjoint parameter $w$, and is expressed as 
\begin{equation}\label{twnrtm}
T_{w} = 8c_B \sin(\pi |\lambda_B|) ~ \frac{1+e^{\i \pi|\lambda_B|}\frac{A_{NR}}{p^{2|\lambda_B|}}}{1-e^{\i \pi|\lambda_B|}\frac{A_{NR}}{p^{2|\lambda_B|}}} \ , ~ ~ ~\text{with},~ ~ A_{NR} = - \frac{1}{w} \Big(\frac{2}{R}\Big)^{2|\lambda_B|}\frac{\Gamma(1+|\lambda_B|)}{1-|\lambda_B|} \ .  
\end{equation}
In this equation, $p$ represents the characteristic non-relativistic momentum scale (like in the scattering of particles with momentum $p$), defined by the near-threshold limit as $$\sqrt{s} = 2c_B+ \frac{p^2}{c_B} \ . $$ 
The quantity $wR^{2|\lambda_B|}$ that appears in \eqref{twnrtm}, is the same quantity that appears in the ACB boundary condition \eqref{acbbc}. 

\vspace{0.15cm} 

Now, when the self-adjoint parameter vanishes, (i.e., $w=0$), it is easy to find that $A_{NR}\to \infty$. In this case, the $T_w$ term of the non-relativistic $S$-matrix (obtained from solving the corresponding Schrodinger problem described above) reduces to 
\begin{equation}
T_{w=0} = - 8c_B \sin(\pi |\lambda_B|) \ . 
\end{equation}
On the other hand, when $w=\infty$, we find that $A_{NR}=0$, so the $T_w$ term simpifies to 
\begin{equation}
T_{w=\infty} = 8c_B \sin(\pi |\lambda_B|) \ . 
\end{equation}

\vspace{0.15cm} 

Thus, we observe that $T_w$ changes  sign when switching from one boundary condition $w=0$ to the other $w=\infty$. This behaviour is consistent with the relative sign change observed in the non-relativistic limit of the full $S$-matrix \eqref{singletfersmatrixsim}. 

\vspace{0.15cm} 

This observation suggests that the non-relativistic Schrodinger equation with the self-adjoint extension parameter $w=\infty$ in the ACB boundary condition \eqref{acbbc} must reproduce the non-relativistic limit of the full $S$-matrix in \eqref{higgsedsmatthreshold}. Hence, this Schrodinger problem with $w=\infty$ is expected to provide a detailed and an accurate description of all aspects of the dynamics of $W$, $\overline{W}$ in the non-relativistic limit. In particular, this schrodinger problem should also incorporate the `near threshold bound state' that describes the $Z$ boson, even though it does not manifest as a pole in the $S$-matrix described above in \eqref{higgsedsmatthreshold}.



\vspace{0.5cm} 

\section{Discussions}\label{sec5}
To summarize, in this paper, we have explored the characteristics of the $Z_\mu$ boson in the higgsed phase of the critical boson theory, which involves fundamental matter coupled to Chern-Simons gauge fields in three spacetime dimensions. We have resolved a key puzzle regarding the characteristics of the $Z$-boson, as well as its dual description in fermionic theory, to which it is mapped under the boson-fermion duality. Working within the unitary gauge \eqref{unitarygaugevev}, we have shown that the $Z$ boson is dual to the $U(1)$ current operator \eqref{u1fcxspace} in the fermionic theory. Consequently, the two-point function of the fermionic $U(1)$ current operator maps, under boson-fermion duality, to the two-point function of the $Z$ boson. 
At the classical level, the fact that the mass of the $Z$ boson is twice the mass of the $W$ boson \cite{Choudhury:2018iwf} suggests that the $Z$ boson can be interpreted as a two-particle bound state of the $W\overline{W}$  at threshold. Since, under Bose-Fermi duality, the elementary fermions correspond to the $W$ bosons, duality implies that $Z$ can plausibly be viewed as a two-particle bound state at the threshold energy of a fermion and an anti-fermion on the dual side. When interactions are taken into account, we find that in the strict large $N$ 't Hooft limit, the $Z$-boson which is classically a stable state in the bosonic theory, evolves into an unstable resonance with a distinctive feature. 

\vspace{0.15cm} 

The spectral densities \eqref{rotrok} associated with the $Z$ boson two-point function exhibit a two-particle $W\overline{W}$ cut starting at $m^2=4c_B^2$. The corresponding spectral width is given by $w=e^{-1/|\lambda_B|}$, which indicates that as the 't Hooft coupling parameter $|\lambda_B|$ approaches zero, the width becomes non-perturbatively small. Consequently, the lifetime of the $Z$ boson grows without bound in the zero coupling limit $|\lambda_B|\to 0$, indicating that the $Z$ boson becomes classically stable \footnote{In other words, considering the decay process $Z_\mu \to W_\alpha +\overline{W}_\beta$ with an interaction of the form $\epsilon_{\mu\alpha\beta}Z^\mu \overline{W}^\beta W^\alpha$ (an interaction term of this kind is present in the action \eqref{hpactioncl} for the critical boson theory in the higgsed phase), it is easy to check that at tree level in (2+1)d, in the rest frame of $Z_\mu$ the decay width is proportioinal to $\Gamma\propto (m_Z^2-4m_W^2)$. Thus, using the classical mass relation \eqref{clmassrels}, it is evident that onshell, the decay width for this decay process vanishes at tree level.\label{fndwz}}.  This is typical of weakly coupled unstable states, which turn into stable states when interactions are turned off.  What is even more intriguing, however, is that rather than taking the usual Lorentzian distribution that one might expect, the spectral density functions instead exhibit a power-law behavior. 

\vspace{0.15cm} 

We have also found that the integrals of the spectral densitites over the full range of $m^2$ diverge. This divergence may lead to a mild puzzle. From the classical action and the equal-time commutation relations of the $Z$ bosons, one would expect the spectral density function to be normalizable, i.e., $$\int_{0}^{\infty}\rho(m^2)dm^2=1 \ . $$ The resolution to this puzzle likely involes renormalization. For non-zero $\lambda_B$, the $Z_\mu$ field that appears on the left hand side of the two-point correlation function \eqref{2ptpred} should presumably be understood as a renormalized operator, while the fields in the canonical commutation relations correspond to the bare field operators. Further exploration of this distinction would be an interesting avenue for future work.

\vspace{0.15cm} 

There are several potential directions for future exploration. One such interesting direction would be to compute the two-point functions of the $Z$ boson directly in the higgsed phase of the critical boson theory coupled to Chern-Simons matter theories without employing the boson-fermion duality, at least in the strict large $N$ 't Hooft limit and to all orders in the 't Hooft coupling. This would allow us to verify whether the results match those used in \eqref{2ptpred}. The conclusions in this paper were obtained based on the assumption of the validity of the Bose-Fermi duality and the application of the corresponding duality map. A direct computation would therefore provide further support to the conjectured Bose-Fermi duality. We leave this issue for future work. 

\vspace{0.15cm}
 
In this paper, we have studied the $Z$ boson by analyzing the two-point function of the $U(1)$ current operator. Another interesting avenue for further study would be to explore the $S$-matrices for $2\to 2$ scattering of $W$ bosons in the higgsed phase of the critical boson theory, to investigate whether spcetral cuts appear and whether they correspond to the $Z$ boson. In Section \ref{dualizedsmatrixwwww}, we have extracted an expression for the singlet channel $S$-matrix of the $W\overline{W}\to W\overline{W}$ scattering process, and have analyzed its properties. Assuming the validity of the 3d boson-fermion correspondence in the non-supersymmetric Chern-Simons matter theories to be true, we feel that the analysis and consequent conclusions made in the Section \ref{dualizedsmatrixwwww} are true. However, a direct computation for the derivation of the corresponding $W$ boson $S$-matrix will be even more satisfying, providing an additional computational evidence for the conjectured 3d Bose-Fermi duality. 
Direct computations of the $2\to 2$ scattering matrices for $W$ bosons in the higgsed phase are currently absent from the literature, partly due to the added computational difficulties arising from the vector nature of the $W$ bosons. Nevertheless, this appears to be feasible task, which we leave for future work. 

\vspace{0.15cm} 

Another potentially interesting direction is to investigate the properties of the $Z$ boson at finite temperature. A preliminary result for the $Z$ boson propagator was suggested in \cite{Mishra:2020wos}. It would be intriguing to investigate how the spectral densities are modified at finite temperature and to understand the fate of the $Z$ boson in the thermal regime. 

\vspace{0.15cm} 

It would also be intriguing to have a careful analysis of the fate of the $Z$ boson when finite $N$ effects are taken into account. A natural progression would be to explore these effects in the non-relativistic limit. For the unitary gauge groups at finite $N$, there are at least three distinct choices for the gauge groups (namely, $SU(N)_{k}$, {\it type II } $U(N)_{k,k}$, and {\it type I} $U(N)_{k, \kappa}$) which have different $\mathcal{O}(1)$ contributions and different choices of the $U(1)$ part of the gauge groups. Hence, the corresponding different dual pairs of non-supersymmetric Chern-Simons matter theories \cite{Minwalla:2020ysu, Minwalla:2022sef} could possibly have different descriptions for the fate of the $Z$ boson. It would be intriguing to gain a clearer understanding of how the properties of the $Z$ boson are modified by the finite $N$ effects across these different theories. 

\section*{Acknowledgements}
The author would like to thank I. Halder, S. Jain, S. Minwalla, and N. Prabhakar for initial collaborations on related works. The author also acknowledges the support from the Yau Center of Southeast University, China. 


\appendix 
 \section{Notation, Conventions, and Duality Maps}\label{appA}
In this paper, we adopt the notations from \cite{Minwalla:2020ysu, Minwalla:2022sef}. For the higgsed phase, we follow the notations and conventions outlined in \cite{Choudhury:2018iwf}. In the bosonic theories, $k_B$ (and similarly $k_F$ for fermionic theories) represents the level of the $SU(N_B)$ (or $SU(N_F)$ for fermions) part of the WZW theory, which is dual to the pure Chern-Simons theory obtained after integrating out the matter fields with a positive mass. The levels $\kappa_B$ and $\kappa_F$, which appear in front of the Chern-Simons action for the gauge fields as in \eqref{cbaction} and \eqref{rfaction}, are referred to as the `renormalized levels'. 
These renormalized levels are related to the above mentioned levels of the corresponding WZW theories through the following relations
\begin{equation}\label{renlevels}
	\kappa_B =k_B +\sgn(k_B)N_B \ , ~ ~ ~ \kappa_F = k_F +\sgn(k_F) N_F \ . 
\end{equation}
We study both the critical boson and regular fermion theories in the strict large $N$ 't Hooft limit, where both ranks and renormalized levels are taken to infinity, while the ratios remain finite. The finite 't Hooft coupling parameters for the bosonic and fermionic theories are defined as 
\begin{equation}\label{defthooftbf}
	\lambda_B = \frac{N_B}{\kappa_B} \ ,  ~ ~  ~ ~  ~ \lambda_F = \frac{N_F}{\kappa_F} \ . 
\end{equation} 
Bose-Fermi duality, which is a form of strong-weak duality, maps the 't Hooft parameters to each other as follows
\begin{equation}
	\lambda_F = \lambda_B - \sgn(\lambda_B)  \ .
\end{equation}
The allowed ranges of the coupling parameters are
\begin{equation}
 0\leq |\lambda_B| , |\lambda_F| \leq 1 \ . 
\end{equation}
The conjectured duality map between the renormalized levels and ranks of the respective gauge groups, is given by 
\begin{equation}
	N_B =|\kappa_F|-N_F \  , ~ ~ ~ ~ \kappa_B = - \kappa_F \ , 
\end{equation}
which can also be rewritten in the following equivalent form 
\begin{equation}
	N_B=|k_B| \ , ~ ~ ~ k_B= - \sgn(k_F) N_F \ . 
\end{equation}
The quasi-fermionic theories are also characterized by an additonal mass parameter. The mass parameter $m_F$ for the regular fermion theory and the mass parameter $m_B^{\text{cri}}$ for the critical boson theory are conjectured to be related to each other in the large $N$ limit by the following equation
\begin{equation}
	m_F =-\lambda_B m_B^{\text{cri}} \ . 
\end{equation}

\vspace{0.15cm} 

For an Euclidean $3$-vector $a^\mu\equiv (a^1, a^2, a^3)$, the corresponding lightcone components are defined in the usual way as 
\begin{equation}
	a^{\pm} = \frac{1}{\sqrt{2}}(a^1\pm \i a^2) \ . 
\end{equation}
In our conventions, the non-zero components of the Euclidean metric in the lightcone coordinates are given by $\delta_{+-}=\delta_{-+}=\delta_{33}=1$. Also, the conventions for the normalization of the totally antisymmetric Levi-Civita tensor in the Euclidean signature is $\epsilon_{123}=\epsilon^{123}=+1$. The corresponding non-zero component of the Levi-Civita tensor in the lightcone coordinates is given by $\epsilon_{+-3}=\epsilon^{-+3}=\i $.

\section{A brief overview of the critical boson theory in the Higgsed phase}
In this appendix, we summarize some of the relevant details of the critical boson theory \cite{Choudhury:2018iwf}, for the analysis in this paper. The critical boson theory is defined by the Lagrangian
\begin{equation}\label{cbaction}
	S_{\text{CB}}[A, \phi] = \frac{\i \kappa_B}{4\pi}\int d^3x ~ \epsilon^{\alpha\beta\gamma}~\text{Tr}\Big(A_\alpha\partial_{\beta}A_{\delta} -\frac{2\i}{3}A_\alpha A_\beta A_\delta \Big) + \int d^3x ~\bigg(\big(D_\mu\bar{\phi} \big)\big(D^\mu \phi\big) + \sigma\Big(\bar{\phi}\phi + \frac{N_B}{4\pi}m_B^{\text{cri}} \Big) \bigg) \ , 
\end{equation}
where, $A_\alpha$ is the Chern-Simons gauge field corresponding to the $SU(N_B)_{k_B}$ gauge group. $\phi$ is a complex scalar field transforming in the fundamental representation of the associated $SU(N_B)_{k_B}$ Chern-Simons gauge group. The field $\sigma$ is a gauge-singlet, auxiliary field that serves as a Lagrange multiplier. This can be interpreted as a Hubbard-Stratonovich field mediating a $\phi^4$ interaction between the bosons. When the coupling of $\phi^4$ is taken to be very large, the quadratic term for the Hubbard-Stratonovich field vanishes, effectively converting it into a Lagrangian multiplier. The covariant derivative is defined by 
\begin{equation}\label{cvd} 
	D_\mu \phi = \partial_\mu \phi - \i A_\mu \phi \ , ~ ~ ~ ~ ~ D_\mu \bar{\phi} = \partial_\mu \bar{\phi} + \i \bar{\phi} A_\mu \ . 
\end{equation}
The mass parameter $m_B^{\text{cri}}$ appearing in \eqref{cbaction} characterizes the critical boson theories. Its sign determines the phase of the scalar bosons: a positive sign corresponds to the unhiggsed phase, while a negative sign indicates the higgsed phase.  

\vspace{0.15cm} 

For the purposes of this paper, we focus on the theory in \eqref{cbaction}, with particular emphasis on the case where $m_B^\text{cri}<0$ which characterizes the higgsed phase. The scalar field acquires a vacuum expectation value (vev) determined by $m_B^{\text{cri}}$, as evident from the equation of motion for the auxiliary field $\sigma$, 
\begin{equation}\label{sigmeom}
	\bar{\phi}_i\phi^i = -\frac{N_B}{4\pi}m_B^{\text{cri}} \ , 
\end{equation}
where, in the LHS of the above equation \eqref{sigmeom}, the color index $i$ is summed over the all values from $i =1, 2, \cdots, N_B$. To study the higgsed phase, it is convenient to work in the unitary gauge, in which the scalar field is chosen to be real and given by
\begin{equation}\label{ugphivalue}
	\phi^i = v\sqrt{|\kappa_B|}  ~\delta^{i N_B} = \delta^{i N_B} \sqrt{\frac{N_B}{4\pi}|m_B^\text{cri}|} \ ,
\end{equation}
where, the second equality in the above expression \eqref{ugphivalue} follows from the equation of motion of the auxiliary filed $\sigma$. The real parameter $v$ appearing in \eqref{ugphivalue} characterizes the vacuum expectation value of the scalar field in the unitary gauge and, and with this choice the scalar field $\phi$ is completely determined. The Higgs mechanism breaks the original $SU(N_B)$ gauge symmetry group to an unbroken subgroup of $SU(N_B-1)$. 

\vspace{0.15cm} 

Due to the symmetry breaking of the gauge group, the matter degrees of freedom are transferred to the gauge field. As the Chern-Simons gauge fields are non-dynamical, the components of the gauge fields which transformed as adjoints under the original gauge group $SU(N_B)$ and no longer transforms as adjoint under the unbroken gauge group $SU(N_B-1)$, becomes massive and the number of all these massive components adds up to $2(N_B-1)+1=2N_B-1$. As expected, this is equal to the total number of independent real components of the complex scalar field $\phi^i$ once the equation of motion for the auxiliary field $\sigma$ in \eqref{sigmeom} is also taken into account. The equation of motion for $\sigma$ in \eqref{sigmeom} freezes the modulus of the complex scalar field. The condensation thus effectively transforms the degrees of freedom from a spin zero scalar to a spin $\pm 1$ vector. 

\vspace{0.15cm} 

Schematically, the breaking of the gauge group takes the following form: 
\begin{equation}\label{suntosunmo}
\mathbf{A}_{N_B\times N_B} \longrightarrow \left(
\mbox{\large$
	\renewcommand{\arraystretch}{1.2}
	\begin{array}{c|c}
		\mathbf{\tilde{A}}_{(N_B-1)\times (N_B-1)} & \mathbf{W}_{(N_B-1)\times 1} \\
		\hline
		\mathbf{\overline{W}}_{1\times (N_B-1)} & Z \\
	\end{array}
	$}
\right) \ , 
\end{equation} 
where, the boldface letters are the matrix representation of the corresponding fields. Since, the Chern-Simons gauge fields are in the adjoint representation of the corresponding gauge group, explicit color indices of the gauge field is labelled as  $(A_\mu)^i_{~j}$, where, $i$ is the fundamental index and $j$ is the anti-fundamental index. As the original gauge group is $SU(N_B)$, both fundamental and anti-fundamental indices of the unbroken gauge group run over the values $i, j=1, 2, \cdots, N_B$. 
More precisely, in the components forms, the $SU(N_B)$ gauge fields can be split in the following way to generate $SU(N_B-1)$ gauge fields \cite{Choudhury:2018iwf}:
\begin{equation}
	\big(A_{\mu}\big)^i_{~j} = \begin{pmatrix}
		\big(\tilde{A}_{\mu}\big)^a_{~b} - \frac{\delta^a_{b}}{N_B-1}Z_\mu & & &  \frac{1}{\sqrt{\kappa_B}}W_{\mu}^a \\ \\
		\frac{1}{\sqrt{\kappa_B}}\overline{W}_{\mu b} & &  &  Z_\mu 
	\end{pmatrix} \ , 
\end{equation}
where, in the above equation, the Latin indices $a, b$ are the gauge indices corresponding to the unbroken $SU(N_B-1)$ gauge group, and it runs over the values $1,2, \cdots, N_B-1$. The $W_\mu$ boson transforms in the fundamental representation of the unbroken gauge group and its complex conjugate transforms in the anti-fundamental representation of the unbroken gauge group. The $Z_\mu$ boson is color-singlet under the unbroken $SU(N_B-1)$ subgroup of the original gauge symmetry group $SU(N_B)$. 

\vspace{0.15cm} 

The classical action of the critical boson theory in the higgsed phase, in the strict large $N$ limit (the order one and the subleading factors drop out in the strict large $N$ limit), is given by 
\begin{equation}\label{hpactioncl}
	\begin{split} 
	S_{H}[\tilde{A}, W, Z] & = \frac{\i \kappa_B}{4\pi}\int d^3x~ \epsilon^{\alpha \beta \gamma}\text{Tr}\Big(\tilde{A}_\alpha \partial_{\beta}\tilde{A}_\gamma -\frac{2\i}{3} \tilde{A}_\alpha \tilde{A}_\beta \tilde{A}_\gamma  \Big)\\ 
	&  + \frac{\i}{2\pi}\int d^3x ~\Big(\epsilon^{\mu\nu\rho} \overline{W}_\mu \partial_\nu W_\rho + \sgn(\kappa_B)2\pi v^2 \overline{W}_\mu W^\mu \Big) \\
	&  + \frac{\i \kappa_B }{4\pi}\int d^3x ~\Big(\epsilon^{\mu\nu\rho} Z_\mu \partial_\nu Z_\rho + \sgn(\kappa_B) 4\pi  v^2 Z_\mu Z^\mu \Big) \\
	& +  \frac{1}{2\pi}\int d^3x ~\epsilon^{\mu\nu\rho} \overline{W}_\mu (\tilde{A}_\nu-Z_\nu) W_\rho \ . 
	\end{split} 
\end{equation}
This action shows that classically the mass of the $Z$ boson is twice that the mass of the $W$ boson, as highlighted in \eqref{clmassrels}. 

\section{Summary of fermionic results for the two-point correlation function of the conserved current}\label{apprf}
In this appendix, we provide a summary of key results from the regular fermion theory, which under boson-fermion duality, maps to the critical boson theory  in both its unhiggsed and higgsed phases. The classical action of the regular fermionic theory coupled with $U(N_F)$ Chern-Simons gauge fields in the large-$N_F$ limit (where we focus on the strict large-$N$ limit and omit the effects of $U(1)$ factors), is given by 
\begin{equation}\label{rfaction}
	S_{RF}[\psi, A] = \frac{\i \kappa_F}{4\pi}\int d^3x~ \epsilon^{\alpha \beta \gamma}\text{Tr}\Big(A_\alpha \partial_{\beta}A_\gamma -\frac{2\i}{3} A_\alpha A_\beta A_\gamma  \Big) + \int d^3x ~ \overline{\psi} \Big( \gamma^\mu (\partial_\mu -\i A_\mu ) + m_F \Big)\psi \ . 
\end{equation}
Here, $m_F$ represents the bare mass of the fermion, and the gauge field $A_\mu$ transforms under the adjoint representation of the $U(N_F)$ gauge group. 

\vspace{0.15cm} 

Since the gauge invariant two-point correlation functions of the $U(1)$ currents match on both sides of the duality, we focus on the two-point correlation function of the corresponding $U(1)$ current in the regular fermionic theory. The conserved current of the regular fermionic theory is given by 
\begin{equation}\label{jmuf}
	J_{\mu}^F = \i \overline{\psi}_i \gamma_\mu \psi^i \ . 
\end{equation}

\subsection{Known results}
The momentum space two-point correlation function
\begin{equation} 
 \langle J_\mu^F(p) J_\nu^F(q)\rangle = \mathcal{G}_{\mu\nu}^F(q) ~ (2\pi)^3 ~\delta^{(3)}(p+q) \ , 
 \end{equation} 
 of the current operator \eqref{jmuf} in the large-$N$ (planar) 't Hooft limit, in the special choice of the light-cone momenta $q_{\pm}=0$ (and in the lightcone gauge $A_{-}=0$) is known in the literature \cite{Gur-Ari:2015pca, Mishra:2020wos} for the $(-+)$ component. The result is summarized as 
 \begin{equation}\label{gfmpq3} 
 	\mathcal{G}^F_{-+}(q_3) = \frac{\i N_F q_3}{16\pi \lambda_F} \Big(1+\frac{2\i c_F}{q_3}\sgn(m_F) \Big)^2 \bigg(\exp\Big(2\i \lambda_F \tan^{-1}\big(\frac{q_3}{2c_F}\big)\Big) -1  \bigg) - \frac{N_F c_F}{4\pi} \ . 
 \end{equation}
 Here, the quantiy $c_F$ is the large-$N_F$ exact pole mass of the regular fermion determined by solving the gap equation, and is given by \cite{Jain:2014nza}
 \begin{equation}\label{cfpolemass}
 	c_F = \frac{m_F}{\sgn(m_F)-\lambda_F} \ .
 \end{equation} 	
 We follow the convention where, the pole mass $c_F$ (and similarly $c_B$ for bosons) is always positive. The expression for $\mathcal{G}^F_{-+}(q_3)$ does not exhibit a definite even/odd symmetry under the following transformation $q_3\to -q_3$. However, it can always be expressed as a sum of  the even and odd components:
 \begin{equation}
 	\mathcal{G}^{F}_{-+}(q_3) = ~  \mathcal{G}^{F, \text{even}}_{-+}(q_3) ~ + ~  \mathcal{G}^{F, \text{odd}}_{-+}(q_3) \ . 
 \end{equation}
 This decomposition is useful because, as we will see below, $\mathcal{G}^{F, \text{even}}_{-+}(q_3)$ corresponds to a specific component of the covariant structure $\mathcal{G}^{F, \text{even}}_{\mu\nu}(q)$ containing the even-tensor structure $\Big(\frac{q^2\delta_{\mu\nu}-q_\mu q_\nu}{|q|}\Big)$. We will refer to this as the even component. Similarly, $\mathcal{G}^{F, \text{odd}}_{-+}(q_3)$ is a component of $\mathcal{G}^{F, \text{odd}}_{\mu\nu}(q)$ containing the odd-tensor structure $\epsilon_{\mu\nu\rho}q^\rho$. Although it may seem unconventional, we will refer to this as the odd component. The procedure for extracting these components will be outlined below.  
 \subsubsection*{Even component}
 The even component is obtained by considering the part of the result \eqref{gfmpq3} that satisfies the following condition
 \begin{equation}
 	\mathcal{G}^{F, \text{even}}_{-+}(-q_3) =  \mathcal{G}^{F, \text{even}}_{-+}(q_3) \ .
 \end{equation}
It can be shown, after some straightforward algebra, that the $q_3$-even part is given by 
 \begin{equation}\label{gfeven}
 \begin{split} 
 	\mathcal{G}^{F, \text{even}}_{-+}(q_3)   = & ~  \bigg[-\frac{N_F}{16\pi \lambda_F}\Big(1-\frac{4c_F^2}{q_3^2}\Big) \sin\bigg(2\lambda_F \tan^{-1}\Big(\frac{|q_3|}{2c_F}\Big)\bigg) \\
 	& - \frac{N_Fc_F}{4\pi \lambda_F|q_3|}\sgn(m_F)  \bigg(\cos\bigg(2\lambda_F \tan^{-1}\Big(\frac{|q_3|}{2c_F}\Big)\bigg)-1\bigg) -\frac{N_Fc_F}{4\pi |q_3|} \bigg]|q_3|  \ . 
 \end{split} 
 \end{equation}
 Before proceeding, we should note that for a fixed choice of the sign of $\sgn(m_F\lambda_F)$, the function in \eqref{gfeven} is an even function of $\lambda_B$. 
 \subsubsection*{Odd component}
 Similarly, the odd component is obtained by considering the part of \eqref{gfmpq3} that satisfies
 \begin{equation}
 	\mathcal{G}^{F, \text{odd}}_{-+}(-q_3) = -  \mathcal{G}^{F, \text{odd}}_{-+}(q_3) \ .
 \end{equation}
 The corresponding result for the $q_3$-odd part is given by 
 \begin{equation}\label{gfodd}
 	\begin{split} 
 		\mathcal{G}^{F, \text{odd}}_{-+}(q_3)  = &  ~- \i  \bigg[-\frac{N_F}{16\pi \lambda_F}\Big(1-\frac{4c_F^2}{q_3^2}\Big) \bigg(\cos\bigg(2\lambda_F \tan^{-1}\Big(\frac{|q_3|}{2c_F}\Big)\bigg)  -1\bigg) \\
 		& + \frac{N_Fc_F}{4\pi \lambda_F|q_3|}\sgn(m_F) \sin\bigg(2\lambda_F \tan^{-1}\Big(\frac{|q_3|}{2c_F}\Big)\bigg) \bigg]q_3 \ . 
 	\end{split} 
 \end{equation}
 It is noteworthy to mention that for a fixed choice of the sign of $\sgn(m_F\lambda_F)$, the function in \eqref{gfodd} is an odd function of $\lambda_F$. 
 
 \subsection{Covariant generalization}
 The two-point current correlation function presented in \eqref{gfmpq3}, was originally computed in a specific lightcone kinematical setup, where $q_{+}=q_{-}=0$ and $q_3$ is finite. To derive a covariant generalization of this result, we can utilize the covariance under $SO(3)$ rotations in momentum space. By replacing $|q_3|$ with $|q|\equiv+\sqrt{q^2}=+\sqrt{q_1^2+q_2^2+q_3^2}$, and noting that $\epsilon_{-+3}=-\i$, the result in \eqref{gfmpq3} can be generalized to the following covariant form 
 \begin{equation}\label{gmunufcov}
 	\mathcal{G}_{\mu\nu}^{F}(q) = f_\text{e}(|q|) \bigg(\frac{q^2\delta_{\mu\nu}-q_\mu q_\nu}{|q|}\bigg) +  f_\text{o}(|q|) \epsilon_{\mu\nu\rho} q^\rho \ . 
 \end{equation}
 The even component, $f_\text{e}(|q|)$, is the covariantized version of the expression in \eqref{gfeven} and is given by 
 \begin{equation}
 	\begin{split}
 		f_\text{e}(|q|)  = &  -\frac{N_F}{16\pi \lambda_F}\Big(1-\frac{4c_F^2}{q^2}\Big) \sin\bigg(2\lambda_F \tan^{-1}\Big(\frac{|q|}{2c_F}\Big)\bigg) \\
 		& - \frac{N_Fc_F}{4\pi \lambda_F|q|}\sgn(m_F)  \bigg(\cos\bigg(2\lambda_F \tan^{-1}\Big(\frac{|q|}{2c_F}\Big)\bigg)-1\bigg) -\frac{N_Fc_F}{4\pi |q|}  \ . 
 	\end{split}
 \end{equation}
In the special case where $(\mu\nu)=(-+)$ and with the lightcone kinematical restriction $q_{\pm}=0$, the even index structure $$\Big(\frac{q^2\delta_{\mu\nu}-q_\mu q_\nu}{|q|}\Big) \ , $$ simplifies to $|q_3|$. Before proceeding, we should note that for a specific choice of the value of $\sgn(m_F\lambda_F)$ (i.e., either $+1$ or $-1$), 
\begin{equation}
	f_{e}(|q|, \lambda_B) = f_{e}(|q|, -\lambda_B) = f_{e}(|q|, |\lambda_B|) \ . 
\end{equation}
In other words, for a specific choice of the phase determining parameter $\sgn(m_F\lambda_F)$, $f_\text{e}(|q|) $ is an even function of $\lambda_B$. 

\vspace{0.15cm} 

The coefficient function $f_\text{o}(|q|)$, which corresponds to the odd term in \eqref{gmunufcov}, is the covariantized form derived from \eqref{gfodd} and is expressed as 
\begin{equation}\label{}
	\begin{split} 
		f_\text{o}(|q|)  = &  -\frac{N_F}{16\pi \lambda_F}\Big(1-\frac{4c_F^2}{q^2}\Big) \bigg(\cos\bigg(2\lambda_F \tan^{-1}\Big(\frac{|q|}{2c_F}\Big)\bigg)  -1\bigg) \\
		& + \frac{N_Fc_F}{4\pi \lambda_F|q|}\sgn(m_F) \sin\bigg(2\lambda_F \tan^{-1}\Big(\frac{|q|}{2c_F}\Big)\bigg) \ . 
	\end{split} 
\end{equation}
As expected, for the particular choice $(\mu\nu)=(-+)$, the odd tensor structure $\epsilon_{\mu\nu\rho} q^\rho$ reduces to $-\i q_3$. As before, we should mention that, for a fixed choice of the value of $\sgn(m_F\lambda_F)$, 
 \begin{equation}
 	f_{o}(|q|, \lambda_B) = - f_{o}(|q|, -\lambda_B) = \sgn(\lambda_B) f_{o}(|q|, |\lambda_B|) \ . 
 \end{equation}
 In other words, for a specific choice of the phase determining parameter $\sgn(m_F\lambda_F)$, $f_\text{o}(|q|) $ is an odd function of $\lambda_B$.

\subsection{Prediction for the Higgsed phase}
 The result in \eqref{gmunufcov} is valid for both possible signs of $m_F$. However, since we are specifically interested in the $Z$ boson in the higgsed phase of the bosonic theory, we consider the case where $\sgn(m_F)= - \sgn(\lambda_F)$ \cite{Choudhury:2018iwf}. This corresponds to the duallity between the regular fermion theory and the critical boson theory in the higgsed phase. To rewrite the result \eqref{gfmpq3} in a more convenient form, we apply the following trigonometric identity
 \begin{equation}\label{exptopowconvform}
 	\exp\Big(2\i \alpha \tan^{-1}(x)\Big) = \bigg(\frac{1+\i x}{1-\i x}\bigg)^{\alpha} \ . 
 \end{equation}
 To obtain a prediction for the two-point function of the  conserved current operator in the higgsed phase, we apply the proposed Bose-Fermi duality map \cite{Choudhury:2018iwf}, which relates the fermionic and bosonic parameters as
 \begin{equation}
 	c_F = c_B, ~ ~ \kappa_F = - \kappa_B, ~ ~ \lambda_F = \lambda_B-\sgn(\lambda_B) \ . 
 \end{equation}
 As mentioned in \eqref{defthooftbf}, the 't Hooft coupling parameters in the fermionic and in the bosonic theory are defined by 
 \begin{equation}
 	\lambda_F=\frac{N_F}{\kappa_F} \ , ~ ~ ~ \lambda_B = \frac{N_B}{\kappa_B} \ . 
 \end{equation}
 Additionally, using the fact that $\sgn(\lambda_F)=-\sgn(\lambda_B)$, we impose the following condition 
 \begin{equation}
 	\sgn(m_F) = \sgn(\lambda_B) \ , 
 \end{equation}
which is valid for the correspondence between the fermionic theory and the higgsed phase of the bosons. Under this Bose-Fermi duality, we find that the predicted two-point function for the $U(1)$ current in the higgsed phase of the bosons is given by 
  \begin{equation}\label{gmunuh}
 	\mathcal{G}_{\mu\nu}^{H}(q) = h_\text{e}(|q|) \bigg(\frac{q^2\delta_{\mu\nu}-q_\mu q_\nu}{|q|}\bigg) +  h_\text{o}(|q|) \epsilon_{\mu\nu\rho} q^\rho \ ,
 \end{equation}
where the coefficients for the even and odd components are given by 
\begin{equation}\label{pepohiggspred}
	\begin{split} 
		h_\text{e}(|q|)  = & ~  \frac{\i N_B}{32\pi |\lambda_B|q^2} \bigg[\frac{(2c_B-i|q|)^4}{(q^2+4c_B^2)}\Big(\frac{2c_B+i|q|}{2c_B-i|q|}\Big)^{|\lambda_B|}-\frac{(2c_B+i|q|)^4}{(q^2+4c_B^2)}\Big(\frac{2c_B-i|q|}{2c_B+i|q|}\Big)^{|\lambda_B|}\bigg] \\
		& - \Big(|\kappa_B|-\frac{N_B}{2}\Big)\frac{c_B}{2\pi |q|} \  ,  \\ 
		h_\text{o}(|q|)  = & -  \frac{N_B}{32\pi \lambda_B q^2} \bigg[\frac{(2c_B-i|q|)^4}{(q^2+4c_B^2)}\Big(\frac{2c_B+i|q|}{2c_B-i|q|}\Big)^{|\lambda_B|}+\frac{(2c_B+i|q|)^4}{(q^2+4c_B^2)}\Big(\frac{2c_B-i|q|}{2c_B+i|q|}\Big)^{|\lambda_B|}\bigg] \\ 
		&  + \frac{N_B}{16\pi \lambda_B}  \Big(\frac{4c_B^2}{q^2}-1\Big) \ .  
	\end{split} 
\end{equation}
These generalized results \eqref{pepohiggspred} in the covariant form are used in the main text of this paper as in \eqref{pepohiggsztwopt}, to study the two-point function (and the spectral representation) of the $Z$ boson in the higgsed phase. 
\section{A note on the evaluation of the discontinuity and the imaginary part}\label{appdisim}
In this appendix, we briefly review the method of computation of the discontinuity and the imaginary parts of the complex functions across a branch cut. Let us consider a function $f(s)$ of a real variable $s$. We make the following two assumptions for the function $f(s)$:
\begin{itemize}
	\item It admits an analytic continutaion in the complex $s$ plane with a branch cut along the positive real axis starting at a (branch) point $s=s_0>0$ and extending to infinity \footnote{For reasons of physical interest relevant to our paper, here we have chosen the the branch cut to lie along the positive real axis. However, a more general situation can be considered and the same argument will follow through if we place the branch cut starting at a point on the real axis and extending to infinity at an angle $\theta$, where this angle is measured relative to the positive real axis. In that case the principle sheet for the function will be given by the region of the complex plane with the restriction on the argument of $s$  $$\text{arg}(s)\in (\theta, \theta+2\pi)\ ,$$ which excludes the branch cut.}. 
	\item For real $s$, along the positive real axis in the range $s<s_0$, the function $f(s)$ is real. 
\end{itemize}
It is evident that for real value of $s<s_0$, the real function $f(s)$ satisfies the following condition 
\begin{equation}\label{ccrelreal}
	f(s) = \big(f(s^\star)\big)^\star \ , 
\end{equation} 
where, in the above equation, the star indicates the complex conjugate. As $f(s)$ admits an analytic continuation in the complex $s$ plance, it is an analytic function of the complex variable $s$. Since, each side of equation \eqref{ccrelreal} is an analytic function of $s$,  \eqref{ccrelreal} can be analytically continued to the whole complex $s$ plane, which in the component form takes the following form
\begin{equation}\label{reimfs}
	\begin{split}
		\text{Re} f(s) & = \text{Re} f(s^\star) \ , \\
		\text{Im} f(s) & = -  \text{Im} f(s^\star) \ . 
	\end{split}
\end{equation}

\vspace{0.15cm} 

Hence we see that, as a consequence of the continuation outside the real axis, $f(s)$ acquires an imaginary part. In other words, upon approaching the real $s$ axis from above with $s+\i \epsilon$ (and $\epsilon$ being understood to be $0^+$ throughout this paper) or from below with $s-\i \epsilon$, \eqref{reimfs} implies that $f(s)$ has opposite imaginary parts $\text{Im} f(s+\i \epsilon)  = -  \text{Im} f(s-\i \epsilon)$.  Since, there is a branch cut across the positive real axis starting at $s=s_0$, upon crossing the real axis across the branch cut, there is a discontinuity 
\begin{equation}\label{imdiscrel}
	\text{Disc}f(s) = f(s+\i \epsilon) - f(s-\i \epsilon) = 2\i~ \text{Im}f(s+\i \epsilon) \ , 
\end{equation}
which vanishes along the real axis with $s<s_0$. Accordingly, the imaginary part of the function $f(s)$ is alternatively given by its discontinuity upon crossing the real axis across the branch cut, which might be easier to determine in actual calculations. The $\i \epsilon$ prescription indicates that for $s>s_0$, the physical domain of the variable $s$ should be above the cut, at $s+\i \epsilon$. Since the branch cut is placed along the positive real axis, the principal sheet for the function $f(s)$ is defined as the region of the complex $s$ with the restriction 
\begin{equation}
	\text{arg}(s) \in (0,2\pi) \ , 
\end{equation}
excluding the positive real $s$ axis (the branch cut).

\vspace{0.15cm}

\subsection*{Determination of the discontinuity and imaginary part of $z^\alpha$}

We apply the above analysis to determine the discontinuity (and hence the imaginary part) of the following function $z^{\alpha}$ (for non-integer $\alpha$) across a branch cut for a complex variable $z$. For simplicity, we decompose the complex variable $z$ into its real part to be denoted by $x$ and its imaginary part to be labelled by $y$. We first note that when $\alpha$ is a non-integer, this function is multi-valued, in particular with a branch point at $z=0$. To make it an unambiguously single-valued function in the physical sheet, a branch cut is placed along the negative real $x$ axis, starting at $z=0$ and extending to negative infinity. A point across the branch cut on the negative real $x$ axis can be parameterized by introducing the $\i \epsilon$ depending upon whether the point is above cut or below the cut. A point which is at a distance $-x$ (for $x<0$) from the origin just above the cut is parameterized as $x+\i \epsilon$ and if the point is below the cut it is parameterized as $x-\i \epsilon$. The principal sheet for the function $z^\alpha$ is defined as the region of the complex plane with the restriction 
\begin{equation}
	\text{arg}(z)\in (-\pi, \pi) \ , 
\end{equation}
meaning the argument of $z$ is constrained to lie between $-\pi$ and $\pi$, excluding the negative real axis (the branch cut) where the argument would jump discontinuously. Hence, we find that the just above the branch cut on the negative real axis, the function $z^\alpha$ takes the value $|z|^\alpha e^{\i \pi \alpha}$ while just below the branch cut the value it takes is $|z|^\alpha e^{-\i \pi \alpha}$. Thus we find that the discontinuity of the function $z^\alpha$ across the branch cut along the negative real axis (as the cut is approached from above first) is given by 
\begin{equation}
	\begin{split} 
	\text{Disc}(z^\alpha) & = |z|^\alpha e^{\i \pi \alpha} - |z|^\alpha e^{- \i \pi \alpha} \ , \\
	& = 2\i |z|^\alpha \sin(\pi \alpha) \ . 
	\end{split} 
\end{equation}
It follows from the relationship \eqref{imdiscrel} that the imaginary part of the function $z^\alpha$ is given by 
\begin{equation}
	\text{Im} (z^\alpha) = |z|^\alpha \sin(\pi \alpha) \ . 
\end{equation}

\vspace{0.15cm} 

\subsection*{A proof of the relationship \eqref{specimrel}}\label{appproofimrho}

Before concluding this appendix, we provide a proof of the statement \eqref{specimrel} relating the imaginary parts of the coefficient functions of the two-point function of the $Z$ boson with the corresponding spectral density functions. It follows from the definition of the spectral decomposition \eqref{spectraldecdef} (we include the $\i \epsilon$ here to make it precise, and work in the $s$ variable for convenience), i.e., 
\begin{equation}
	\tau(s) = \int dm^2 ~ \frac{\rho(m^2)}{-s+m^2-\i \epsilon} \ , 
\end{equation}
that the imaginary part of the function $\tau(s)$, which is defined to be equal to $\tau(s)-(\tau(s))^\star$ divided by $2\i$, is given by 
\begin{equation}\label{imtausdef}
	\text{Im}(\tau(s)) = \frac{1}{2\i} \int dm^2 ~ \rho(m^2) \bigg(\frac{1}{-s+m^2-\i \epsilon} - \frac{1}{-s+m^2+\i \epsilon} \bigg) \ . 
\end{equation}
We use the fact that a function of the form $(x-\i \epsilon)^{-1}$ for a real variable $x$, should be realized as the principal value (conventionally abbreviated as Pv) $\text{Pv}\big(\frac{1}{x}\big)$ plus an imaginary contribution of the form $\i \pi \delta(x)$. In other words, 
\begin{equation}\label{pvinvep}
	\frac{1}{x-\i\epsilon} = \text{Pv}\Big(\frac{1}{x}\Big) + \i\pi  \delta(x) \ . 
\end{equation}
It is easy to see by substituting \eqref{pvinvep} in the integrand of \eqref{imtausdef}, that the principal value part from both the terms cancel each other, while the imaginary parts containing the $\delta$-function add up. Hence, we find that 
\begin{equation}
	\begin{split}
			\text{Im}(\tau(s)) & = \frac{1}{2\i} \int dm^2 ~ \rho(m^2) ~ \Big(2\pi \i \delta(m^2-s) \Big) \ , \\ 
			& = \pi ~\rho(s) \ . 
	\end{split}
\end{equation}
This completes the proof of the statement \eqref{specimrel}. 

\section{Plots of the spectral density functions}\label{appplots}
In this appendix, we present plots of the spectral density functions for the even and the odd components of the two-point function of the $Z$ boson, with an overall normalization of the fields. The funcitonal forms of the spectral density functions, $\rho_\tau(s)$ and $\rho_\kappa(s)$, are given by eq. \eqref{rotrok}. We focus on the plots of these functions for sufficiently small values of the bosonic `t Hooft coupling parameter, $\lambda_B$, to illustrate how the spectral density functions broaden as $|\lambda_B|$ is increased from zero to an infinitesimal value. This behaviour indicates that the corresponding $Z$-boson becomes unstable for any non-zero value of the bosonic 't Hooft coupling parameter.  

\vspace{0.15cm} 

\begin{figure}[H]
	\centering
	\includegraphics[width=0.85\textwidth]{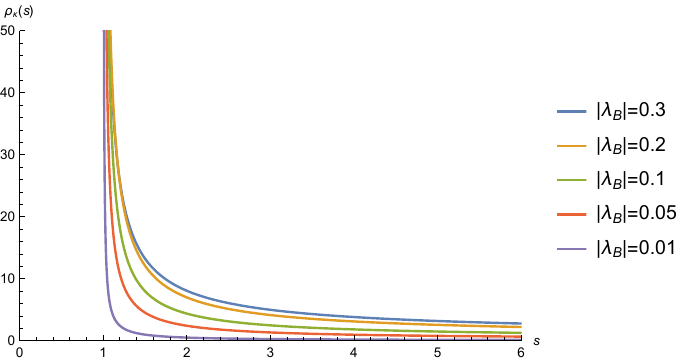}
	\caption{Plots of the spectral density function $\rho_\kappa(s)$ \eqref{rotrok}, corresponding to the odd component of the (rescaled) $Z$-boson two-point function, for several small, fixed values of $|\lambda_B|$.}
	\label{fig:rhokplot}
\end{figure}

To present the plots of the spectral density functions, it is important to note that these plots are scaled relative to one another, rather than being plotted to an absolute scale, in order to highlight the key features. The variable $s$ is plotted along the horizontal axis, with a scaling $$c_B=\frac{1}{2} \ , $$ chosen for convenience. Figures \ref{fig:rhotplot} and \ref{fig:rhokplot} display the spectral density functions $\rho_\tau(s)$ and $\rho_{\kappa}(s)$, respectively, for several small values of $|\lambda_B|$, as shown below: 
\begin{figure}[H]
	\centering
	\includegraphics[width=0.85\textwidth]{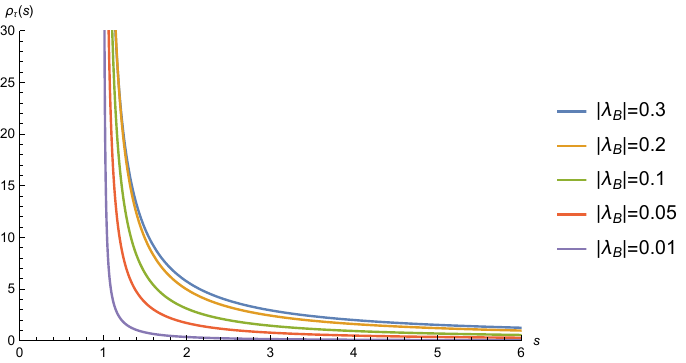}
	\caption{Plots of the spectral density function $\rho_\tau(s)$ given in \eqref{rotrok}, corresponding to the even component of the (rescaled) $Z$-boson two-point function, for several small, fixed values of $|\lambda_B|$.}
	\label{fig:rhotplot}
\end{figure}

\vspace{0.15cm}

The plots in Figures \ref{fig:rhotplot} and \ref{fig:rhokplot} clearly show a spike near the the weak coupling limit, specifically around $|\lambda_B| \approx 0$. This indicates the presence of an approximate pole in the spectrum at $s=4c_B^2$, in the bosonic zero coupling limit. Additionally, both figures \ref{fig:rhotplot} and \ref{fig:rhokplot} reveal that in the weak coupling regime, as $|\lambda_B|$ increases slightly (as an example from $|\lambda_B|=0.01$ to $|\lambda_B|=0.05$), there is a significant broadening of the spectral density functions $\rho_\tau(s)$ and $\rho_\kappa(s)$. This behaviour is reminiscent of our observation in this paper that the $Z$ boson becomes unstable at non-zero values of the `t Hooft coupling, decaying into $W ~\text{and}~\overline{W}$ particles, but with a non-perturbatively large lifetime. 

\vspace{0.15cm}

At first glance, it might seem concerning that the plots of the spectral density functions for different values of $|\lambda_B|$, in Figures \ref{fig:rhotplot} and \ref{fig:rhokplot}, intersect at certain values of the Mandelstam variable $s$. However, this is not an issue, but rather an essential feature for the correct interpretation of these functions. Specifically, such crossings are required to enable a meaningful `area under the curve' interpretation of the integral of the spectral density functions over a specified range of $s$, as outlined in \eqref{normaucperhot}. This interpretation holds even though the integral of the spectral density functions over the entire range of $s$, from zero to infinity, diverge, as discussed in \eqref{phoint}. 

\vspace{0.15cm}

\begin{figure}[H]
	\centering
	\includegraphics[width=0.85\textwidth]{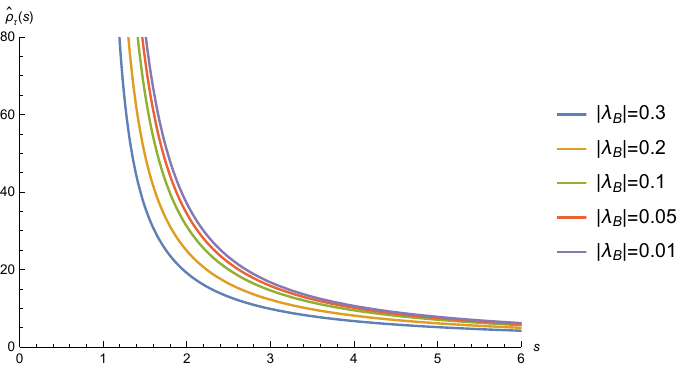}
	\caption{Plots of the spectral density function $\hat{\rho}_\tau(s)$ associated with the even component of the two-point function of $J^H_\alpha$ defined in \eqref{normjz}.}
	\label{fig:tlrhotplot}
\end{figure}

\begin{figure}[H]
	\centering
	\includegraphics[width=0.85\textwidth]{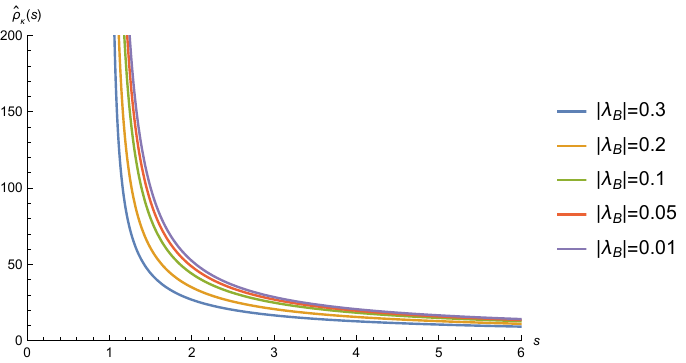}
	\caption{Plots of the spectral density function $\hat{\rho}_\kappa (s)$ associated with the odd component of the two-point function of $J^H_\alpha$ defined in \eqref{normjz}.}
	\label{fig:tlrhokplot}
\end{figure}

This characteristic (of crossing of graphs in Figures \ref{fig:rhotplot} and \ref{fig:rhokplot}) arises from the choice to include an overall normalization factor $\big(\frac{2\pi}{|\kappa_B|}\big)^{1/2}$ in the definition of the normalized current $\tilde{J}_\alpha$, as given in \eqref{normjz}. If this additional normalization factor, which involves an extra term $\frac{1}{\sqrt{|\kappa_B|}}$, had not been considered, the spectral density functions (marked with a caret symbol to distinguish them from $\rho_\tau$ and $\rho_\kappa$ in \eqref{rotrok}), corresponding to the two-point function of the current operator $J_\alpha^H$ would take the following form: 
\begin{equation}
	\hat{\rho}_\tau(s) =\frac{|\kappa_B|}{2\pi} ~ \rho_\tau (s) \ , ~ ~ ~ ~ \hat{\rho}_\kappa(s) =\frac{|\kappa_B|}{2\pi} ~ \rho_\kappa (s) \ . 
\end{equation}
The explicit functional forms in these (unnormalized) spectral density functions can be expressed as 
\begin{equation}\label{rotrokunnorm}
	\begin{split} 
		\hat{\rho}_\tau(s) & = \frac{N_B}{32\pi s^{3/2}} \frac{\sin(\pi|\lambda_B|)}{\pi |\lambda_B|}~ \bigg[ \frac{(s-4c_B^2)^{3-|\lambda_B|}}{(2c_B+\sqrt{s})^{4-2|\lambda_B|}}+\frac{(2c_B+\sqrt{s})^{4-2|\lambda_B|}}{(s-4c_B^2)^{1-|\lambda_B|}}\bigg] \Theta(s-4c_B^2) \  ,  \\ \\
		\hat{\rho}_\kappa(s) & = \frac{N_B}{32\pi s}\frac{\sin(\pi |\lambda_B|)}{\pi|\lambda_B|}~ \bigg[-\frac{(s-4c_B^2)^{3-|\lambda_B|}}{(2c_B+\sqrt{s})^{4-2|\lambda_B|}}+\frac{(2c_B+\sqrt{s})^{4-2|\lambda_B|}}{(s-4c_B^2)^{1-|\lambda_B|}}\bigg]  \Theta(s-4c_B^2)  \ . 
	\end{split} 
\end{equation}
Although these spectral density functions in \eqref{rotrokunnorm} exhibit the standard anyonic factor $$\frac{\sin(\pi|\lambda_B|)}{\pi |\lambda_B|}$$ in the expressions for the two point functions of the $U(1)$ current operators, it is evident from their plots in Figures \ref{fig:tlrhotplot} and \ref{fig:tlrhokplot} that they obsecure key physical features, such as the appearance of the pole at $|\lambda_B|=0$, as well as the spiking behaviour and spectral broadening that occurs when $|\lambda_B|$ is close to zero. This is the reason the additional normalization factor $\sqrt{2\pi/|\kappa_B|}$ was introduced in the definition of the normalized current operator $\tilde{J}_\alpha$ in \eqref{normjz}. The use of this normalization factor in \eqref{normjz} ensures that the spectral density functions are appropriately scaled, allowing for a clearer representations of the subtle features tha emerge as $|\lambda_B|$ approaches zero. 

\vspace{0.15cm}

However, if one were to directly consider the two-point function $\langle Z_\alpha Z_\beta\rangle$, with $Z$ normalized according to the higgsed phase action \eqref{hpactioncl}, without any additional overall normalization factor, the spectral density functions (denoted by a tilde to distinguish them from the other normalization schemes) associated with these two-point functions would be related to those in \eqref{rotrok} through the following expressions
\begin{equation}\label{rhoznorm}
	\tilde{\rho}_\tau(s) =\frac{2\pi}{|\kappa_B|m_Z^2} ~ \rho_\tau (s) \ , ~ ~ ~ ~ \tilde{\rho}_\kappa(s) =\frac{2\pi}{|\kappa_B|m_Z^2} ~ \rho_\kappa (s) \ . 
\end{equation}
In this normalization, the explicit functional forms of the spectral density functions for $\langle Z_\alpha Z_\beta\rangle$ are given by 
\begin{equation}\label{rotroknorm}
	\begin{split} 
		\tilde{\rho}_\tau(s) & = \frac{|\lambda_B|}{16N_B m_Z^2 s^{3/2}} \sin(\pi|\lambda_B|)~ \bigg[ \frac{(s-4c_B^2)^{3-|\lambda_B|}}{(2c_B+\sqrt{s})^{4-2|\lambda_B|}}+\frac{(2c_B+\sqrt{s})^{4-2|\lambda_B|}}{(s-4c_B^2)^{1-|\lambda_B|}}\bigg] \Theta(s-4c_B^2) \  ,  \\ \\
		\tilde{\rho}_\kappa(s) & = \frac{|\lambda_B|}{16N_Bm_Z^2 s}\sin(\pi |\lambda_B|)~ \bigg[-\frac{(s-4c_B^2)^{3-|\lambda_B|}}{(2c_B+\sqrt{s})^{4-2|\lambda_B|}}+\frac{(2c_B+\sqrt{s})^{4-2|\lambda_B|}}{(s-4c_B^2)^{1-|\lambda_B|}}\bigg]  \Theta(s-4c_B^2)  \ . 
	\end{split} 
\end{equation}
The plots of these spectral functions, using the normalization specified in equation \eqref{rhoznorm} and keeping the overall factor of $\frac{1}{16N_B m_Z^2}$ fixed, are provided by the Figures \ref{fig:rhottlplot} and \ref{fig:rhoktlplot} below: 
\begin{figure}[H]
	\centering
	\includegraphics[width=0.85\textwidth]{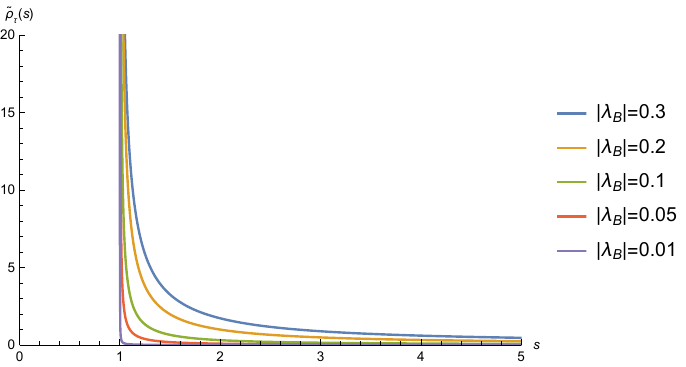}
	\caption{Plots of the spectral density function $\tilde{\rho}_\tau(s)$ associated with the even component of the two-point function of the $Z$-boson.}
	\label{fig:rhottlplot}
\end{figure}

\begin{figure}[H]
	\centering
	\includegraphics[width=0.85\textwidth]{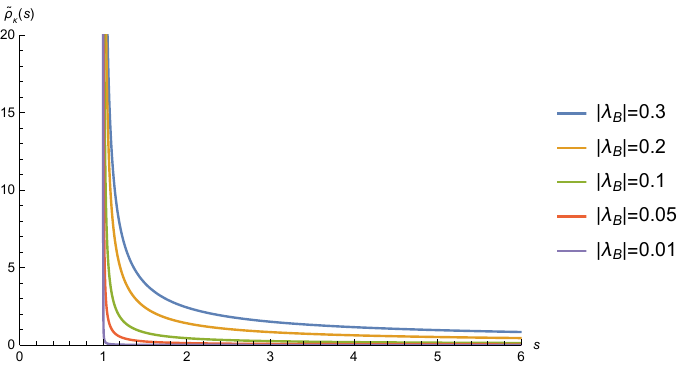}
	\caption{Plots of the spectral density function $\tilde{\rho}_\kappa (s)$ associated with the odd component of the two-point function of the $Z$-boson.}
	\label{fig:rhoktlplot}
\end{figure}

\bibliography{refs}\bibliographystyle{JHEP}

\end{document}